\documentclass[12pt]{article}

\usepackage{amsmath, amsthm, amssymb}
\usepackage{setspace}
\usepackage{multirow}
\usepackage{graphicx}
\usepackage{apacite}
\usepackage{natbib}
\usepackage{textcomp}

\newcommand{\uA}       {\mbox{\boldmath$A$}}

\newcommand{\uB}       {\mbox{\boldmath$B$}}
\newcommand{\ub}       {\mbox{\boldmath$b$}}
\newcommand{\uC}       {\mbox{\boldmath$C$}}

\newcommand{\uD}       {\mbox{\boldmath$D$}}
\newcommand{\uE}       {\mbox{\boldmath$E$}}
\newcommand{\uF}       {\mbox{\boldmath$F$}}
\newcommand{\uG}       {\mbox{\boldmath$G$}}

\newcommand{\uH}       {\mbox{\boldmath$H$}}
\newcommand{\uI}       {\mbox{\boldmath$I$}}

\newcommand{\uQ}       {\mbox{\boldmath$Q$}}
\newcommand{\uR}       {\mbox{\boldmath$R$}}
\newcommand{\uS}       {\mbox{\boldmath$S$}}
\newcommand{\us}       {\mbox{\boldmath$s$}}

\newcommand{\uu}       {\mbox{\boldmath$u$}}

\newcommand{\uV}       {\mbox{\boldmath$V$}}
\newcommand{\uv}       {\mbox{\boldmath$v$}}
\newcommand{\uw}       {\mbox{\boldmath$w$}}
\newcommand{\uX}       {\mbox{\boldmath$X$}}
\newcommand{\ux}       {\mbox{\boldmath$x$}}

\newcommand{\uY}       {\mbox{\boldmath$Y$}}
\newcommand{\uy}       {\mbox{\boldmath$y$}}

\newcommand{\uz}       {\mbox{\boldmath$z$}}

\newcommand{\ualpha}       {\mbox{\boldmath$\alpha$}}
\newcommand{\ubeta}       {\mbox{\boldmath$\beta$}}
\newcommand{\ugamma}       {\mbox{\boldmath$\gamma$}}
\newcommand{\udelta}       {\mbox{\boldmath$\delta$}}

\newcommand{\uepsilon}       {\mbox{\boldmath$\epsilon$}}
\newcommand{\ueta}       {\mbox{\boldmath$\eta$}}
\newcommand{\uEta}       {\mbox{\boldmath$H$}}
\newcommand{\utheta}       {\mbox{\boldmath$\theta$}}

\newcommand{\umu}       {\mbox{\boldmath$\mu$}}

\newcommand{\uSigma}       {\mbox{\boldmath$\Sigma$}}

\newcommand{\uOmega}       {\mbox{\boldmath$\Omega$}}

\newcommand{\uwbar} {\bar{\boldsymbol{\hspace{-2pt}w}}}

\newcommand{\uxi}       {\mbox{\boldmath$\xi$}}
\newcommand{\upi}       {\mbox{\boldmath$\pi$}}

\newcommand{\uzero}       {\mbox{\boldmath$0$}}

\title{Bayesian nonparametric models for spatially indexed data of mixed type}
\author{Georgios Papageorgiou\vspace{0.1cm}\\
Department of Economics, Mathematics and Statistics\\ Birkbeck, University of London, UK \vspace{0.3cm} \\
Sylvia Richardson \vspace{0.1cm}\\
MRC Biostatistics Unit, University of Cambridge, Cambridge, UK \vspace{0.3cm}\\
Nicky Best \vspace{0.1cm}\\
Department of Epidemiology and Biostatistics, Imperial College London, UK\vspace{0.8cm}\\
\emph{Address for correspondence}: Georgios Papageorgiou, Department of Economics,\\ 
      Mathematics and Statistics, Birkbeck, University of London, Malet Street,\\
      London WC1E 7HX, UK\\ 
E-mail: g.papageorgiou@bbk.ac.uk}
\date{}
\begin{document}
\maketitle

\begin{center}
\emph{Abstract}
\end{center}
We develop Bayesian nonparametric models for spatially indexed data of mixed type. 
Our work is motivated by challenges that occur in environmental epidemiology, where the usual presence of 
several confounding variables that exhibit complex interactions and high correlations makes it difficult 
to estimate and understand the effects of risk factors on health outcomes of interest.
The modeling approach we adopt assumes that responses and confounding variables are manifestations 
of continuous latent variables, and uses multivariate Gaussians to jointly model these. Responses and confounding 
variables 
are not treated equally as relevant parameters of the distributions of the 
responses only are modeled in terms of explanatory variables or risk factors. Spatial dependence is introduced by allowing the weights of the 
nonparametric process priors to be location specific, obtained as probit transformations of Gaussian Markov random fields. 
Confounding variables and spatial configuration have a similar role in the model, in that they only influence, 
along with the responses, the allocation probabilities of the areas into the mixture components, thereby  
allowing for flexible adjustment of the effects of observed confounders, while allowing for the possibility of residual 
spatial structure, possibly occurring due to unmeasured or undiscovered spatially varying  factors. 
Aspects of the model are illustrated in simulation studies and an application to a real data set.

\section{Introduction}

In observational studies the task of identifying important predictors for an outcome of interest can be impeded by 
the presence of complex interactions and high correlations among confounding variables. 
Adequately controlling for the effects of such variables can be a challenging task and it may require
the inclusion of main and high order interaction effects in the linear predictor of the model,
while being subject to multicollinearity problems, which
usually lead researchers to select an arbitrary subset of variables to include in the model. 
Hence, the purpose of this article is to propose a general framework, suitable for spatially 
structured data, aiming at inferring the effects of explanatory variables on possibly multivariate 
responses of mixed type, consisting of continuous, count and categorical responses, in the presence of 
confounding variables, also of mixed type, that can exhibit complex interactions and high correlations. 

We consider data observed on the spatial domain, such as point referenced and lattice or
regional data. Although in this article we emphasize regional data, as these are predominant in 
epidemiologic applications which is our specific focus, the presented methods can easily be adapted to accommodate point 
referenced data. In the sequel, we will use subscript $i$ to denote the $i$th region, $i=1,\dots,n$,
of the spatial domain.

Observed data will be classified in three categories. 
With $\uy_i$ we will denote a vector of length $p$ of response variables
observed in area $i$. In our context, these
will be health outcomes, such as numbers of hospitalizations due to different diseases. 
With $\ux_i$ we will denote a collection of explanatory variables or risk factors   
thought to be affecting the response variables. Examples of such variables can
include exposure to air pollution and cigarette smoking. Our interest
is to directly quantify the effects of explanatory variables on the means of the distributions of the responses. Lastly,
with $\uw_i$ we will denote a collection of confounding variables, i.e. variables 
that are thought to have an effect on the distribution of the responses but quantification of their effects is 
not of particular interest. Of interest is only the adjustment for their effects, 
and that is what distinguishes them from the explanatory variables. Examples of  
confounding variables can include area-wise ethnic distributions and exposure to socioeconomic deprivation.

Typically, the adjustment for the effects of confounders $\uw_i$ is made by modeling
the mean parameter of the distribution of responses $\uy_i$ in terms of both confounders $\uw_i$ and 
explanatory variables $\ux_i,$ using the usual regression tool. The regression function can also be obtained indirectly, 
by considering conditional densities of the form $f(\uy_i|\ux_i,\uw_i;\utheta_i^*)$. 
Here we propose to adjust for the effects of $\uw_i$ on the distributions of the responses 
$\uy_i$ by considering joint densities for responses and confounders, $f(\uy_i,\uw_i|\ux_i;\utheta_i)$.
Responses and confounding variables are not treated equally as only parameters describing  
responses are modeled in terms of explanatory variables.  

A benefit of the joint modeling approach is that it frees us
from having to include in the linear predictor main and interaction effects 
of variables $\uw_i$ that are not of particular interest. 
A difficulty that the proposed approach creates is that of having to specify densities for the 
possibly high dimensional vector $(\uy_i,\uw_i)$. To mitigate this difficulty and 
allow for the needed flexibility, we will adopt a Bayesian nonparametric approach.
A further criticism is that the approach classifies covariates,
$(\ux_i, \uw_i),$ as fixed and random purely on the basis of the needs of specific data 
analyses. Hence this classification can change between data analysts depending on their interests,
possibly without any theoretical justification as to why this distinction can be made in the first 
place. See \citet{Mulleretal96} and \citet{MQ10} on issues with considering covariates as random.           

Our modeling approach is related to that of \citet{Mulleretal96}
who jointly modeled continuous data $(\uy_i,\ux_i,\uw_i)$ using mixtures of multivariate normal densities  
and adopted a predictive approach in order to quantify the effects of $\ux_i$ on $\uy_i$. 
Although this is a very general approach, in our context 
quantification of the effects of explanatory variables on responses 
cannot be done by a predictive approach. This is due to the spatial nature of the modeling 
problem that prescribes predictions to be area specific and therefore over a 
restricted range of $\ux_i$. Hence, here we consider densities of the form 
$f(\uy_i,\uw_i|\ux_i;\utheta_i)$ that allow for direct quantification of the effects of 
$\ux_i$ on $\uy_i$ through the regression coefficients. 

Other related modeling approaches include those of
\citet{Shahbaba} and \citet{Hannah}. These authors consider joint models for responses 
and covariates, $f(\uy_i,\ux_i,\uw_i|\utheta^{'})$, as in \citet{Mulleretal96}, but they 
further decompose these densities as conditional and marginal densities:   
$f(\uy_i,\ux_i,\uw_i|\utheta^{'}) = g(\uy_i|\ux_i,\uw_i;\utheta^{'})h(\ux_i,\uw_i|\utheta^{'})$. 
Although this approach performs well in many regression settings,  
it may continue to be problematic in applications in environmental epidemiology 
where confounding variables can exhibit interactions and correlations.
To see this, consider the case of continuous $\ux_i$ and $\uw_i$ modeled by a multivariate Gaussian
density $h(|)$ with unconstrained covariance matrix. The problems caused by correlated and interacting confounding 
variables in models of the form $f(\uy_i|\ux_i,\uw_i;\utheta_i^*)$ will also be present in density $g(|)$ as it 
expresses the mean of $\uy_i$ in terms of 
both explanatory and confounding variables. Here, within a countable mixture 
framework, we examine two possible remedies for this problem. Firstly, 
we will consider mixtures of multivariate Gaussians with covariance matrix
restricted to be diagonal. A potential effect of this restriction is to decompose
the overall dependence among confounding variables into clusters \citep{Henning}
thereby diminishing the multicollinearity problems within density $g(|)$. 
Another potential effect of this restriction, however, is to create many small clusters, 
with a within-cluster regression of $\uy_i$ on $\ux_i$ providing highly variable posterior samples.
Secondly, we will consider densities $g(|)$ which express the mean 
of $\uy_i$ in terms of the explanatory variables only, i.e. densities $g(|)$
with the restriction of regression coefficients corresponding to confounding variables
to be equal to zero. Under these constraints, adjustment for the effects of 
confounding variables is achieved through density $h(|)$.  
           
As indicated earlier, the choice of the priors for $\utheta_i, i=1,\dots,n,$ is crucial in our attempt to 
flexibly adjust for confounder effects, while allowing for spatial dependence among observations at 
nearby areas, potentially occurring due to 
spatially varying unmeasured or undiscovered factors. Specifically, we adopt a nonparametric approach 
by which the area specific prior distributions, $P_i(\utheta)$, are taken to be unknown and modeled using 
dependent nonparametric processes. Starting with the early work of \citet{SNM99}, 
dependent nonparametric processes have become increasingly popular due to the flexibility they 
provide in modeling collections 
of prior distributions, $\{P_1(.),\dots,P_n(.)\},$ the members of which change smoothly with covariates, 
the spatial configuration in our context. Priors $P_i(.)$ corresponding to nearby 
locations can be nearly identical while priors corresponding to areas far apart can be quite different.   
It is this feature of our prior specification that allows for spatial dependence 
among observations at nearby locations. We induce dependence among the members of the collection of 
priors by modeling the $P_i(.)$ as countable discrete mixture distributions with 
weights indexed by $i$: 
$P_i(\utheta) = \sum_{h=1}^{\infty} \pi_{hi} \delta_{\utheta_h}(\utheta)$. 
Here we obtain location-specific weights by utilizing the probit stick breaking processes of 
\citet{RD11} by which the mixture weights are expressed as probit transformations of latent 
Gaussian Markov random fields (GMRFs)
\citep{GMRFbook}. As such, our approach for accounting for potential spatial dependence is related to 
the approach of \citet{FG2002} who considered logistic transformations of GMRFs within a  
finite mixture of Poisson probability mass functions (pmfs) model.  

The density of area $i$ takes the form of a convolution 
$f_{i}(\uy_{i},\uw_{i}|\ux_{i}) = \int f(\uy_{i},\uw_{i}|\ux_{i};\utheta) dP_i(\utheta)$,
where, due to the discreteness of the nonparametric process, density $f_i$ can be expressed as
$f_{i}(\uy_{i},\uw_{i}|\ux_{i}) = \sum_{h=1}^{\infty} \pi_{hi} f(\uy_{i},\uw_{i}|\ux_{i};\utheta_h)$,
where $\pi_{hi}$ are location specific weights.
The resulting mixture formulation provides an effective way of estimating the effects of explanatory 
variables $\ux_i$ on response variables $\uy_i$, while adjusting for the effects of confounding variables.  
To elaborate, adjustment for the effects of confounding variables is achieved by creating clusters
of geographical areas that are similar in terms of the observed values of the confounders. 
With confounding variables having homogeneous values, a  within cluster regression
of $\uy$ on $\ux$ would reflect the true i.e. unconfounded effects of $\ux$ on $\uy$
in that cluster. 

The density of area $i$  can also be expressed as 
$f_{i}(\uy_{i},\uw_{i}|\ux_{i}) = \sum_{h=1}^{\infty} \pi_{hi} g(\uw_i|\utheta_h) f(\uy_{i}|\uw_{i},\ux_{i};\utheta_h)=
\sum_{h=1}^{\infty} \pi_{hi}(\uw_i) f(\uy_{i}|\uw_{i},\ux_{i};\utheta_h) $, where $\pi_{hi}(\uw_i)=\pi_{hi} g(\uw_i|\utheta_h)$. The last 
formulation illustrates how both the regression coefficients and density change with both the spatial locations and vectors $\uw_i$. 
As such, the proposed model is related to the model for density regression described by \citet{DPP}.
Conditional models, 
$f_{i}(\uy_{i}|\uw_{i},\ux_{i}) = \sum_{h=1}^{\infty} \pi_{hi} f(\uy_{i}|\uw_{i},\ux_{i};\utheta_h^{*})$, in contrast, 
allow the density to change with spatial locations only. 
Furthermore, even with relatively simple, e.g. linear, within cluster regression models, 
the mixture formulation allows for complex regression functions to be captured. For instance, the joint 
model $f_{i}(\uy_{i},\uw_{i}|\ux_{i})$ as expressed above, implies that 
$E(\uY_i|\uw_{i},\ux_{i}) = \sum_{h=1}^{\infty} \pi_{hi}(\uw_i) E(\uY_{i}|\uw_{i},\ux_{i};\utheta_h)$.
This method for flexible regression surface estimation was first proposed by \citet{Mulleretal96}. 
The flexibility allowed in the estimation of both the densities and regression surfaces are 
the reasons for which we opted for mixture based nonparametric methods. 
There are of course several other approaches to nonparametric Bayesian 
estimation such as splines, wavelets and neural networks (see \citet{muller2004} for a review).
However, such methods allow only for flexible estimation of regression surfaces.

With the interpretation given above, the proposed model can be thought of as a spatially varying coefficient model. 
Alternatives to the proposed model, for univariate responses, build on the work of \citet{Besag74} and \citet{BK95}. 
\citet{Assuncao} provided several alternative space varying coefficient models suitable for data observed on small areas. 
Further, for multivariate small area responses, such models can be constructed utilizing the methods 
described by \citet{Mardia}, \citet{Jin2005} and \citet{Gelfand2003}. In this paper, we utilize these 
methods to construct a spatially varying coefficient model that can accommodate mixed type responses
as a means of comparison with the proposed mixture based model.  

Our main goal here is to describe general models of the form 
$f_{i}(\uy_{i},\uw_{i}|\ux_{i})= \sum_{h=1}^{\infty} \pi_{hi} f(\uy_{i},\uw_{i}|\ux_{i};\utheta_h)$
were vectors $\uy_i$ and $\uw_i$ can include continuous, count and categorical measurements. 
We jointly model all measurements by assuming that the discrete ones are 
discretized versions of continuous latent variables, and using multivariate Gaussians to jointly 
describe the distributions of observed and latent continuous variables \citep{Muthen}.
As such, our approach is related to the recent work of \citet{DeYoreo} who describe nonparametric mixture models
for binary regression, also utilizing latent variables. 
Models that utilize latent variables, compared to models that assume local independence, 
allow for more flexible clustering by imposing no unnecessary restrictions on the orientations of the 
mixture components. Further, in the context of \citet{DeYoreo}, introduction of latent responses is key to flexibly capturing regression relationships. 

The remainder of this paper is arranged as follows. Section \ref{model} provides a detailed description
of our model formulation. Section \ref{mcmc} provides a brief description of the MCMC algorithm
we have implemented, with most of the technical details deferred to the Appendix. 
Aspects of the model are illustrated in Sections \ref{simulation} and \ref{application} that 
present results from simulation studies and an application to a real dataset which examines the association 
between exposure to air pollution and two birth outcomes. The paper concludes with a brief discussion. 
Samplers for the described models, and some of their special cases, are available in the R package
BNSP \citep{GP}. 

\section{Model specification}\label{model}

The first subsection provides a description of the model formulation for the 
observed data, while the second one provides a description of how the spatial 
configuration is build into the model.

\subsection{Observed data model}

Let $\uy_{i}=(y_{i1},\dots,y_{ip})^T$ denote the vector of mixed type responses observed at location $i$, $i=1,\dots,n$. 
We assume that the elements of $\uy_{i}$ are ordered in the following way: 
the first $p_1$ of them are counts and they are followed by $p_2$ binomial and $p_3$ continuous elements. We jointly model all responses
by assuming that they are manifestations of continuous latent variables denoted by $\uy_{i}^*=(y_{i1}^*,\dots,y_{ip}^*)^T$ \citep{Muthen}. 
In the next few paragraphs we describe the models that connect observed and latent variables. 

Firstly, observed counts and corresponding latent variables are connected through the rule: 
$y_{ik} = t_1(y^*_{ik},\gamma_{ik}) =\sum_{q=0}^{\infty} q I[c_{ik,q-1} < y^*_{ik} < c_{ik,q}]$, $k=1,\dots,p_{1}$.
Here, $I[.]$ denotes the indicator function, $c_{ik,-1} = -\infty$, and for $l \geq 0$, 
$c_{ik,l} = c_{l}(\gamma_{ik}) = \Phi^{-1}\{F(l;\gamma_{ik})\},$ 
where $\Phi(.)$ is the cumulative distribution function (cdf) of 
a standard normal variable, and $F(.;\gamma)$ is the cdf of a Poisson$(\gamma)$ variable. 
It is clear from the definitions of the cut-points that marginally $Y_{ik} \sim \text{Poisson}(\gamma_{ik}), k=1,\dots,p_{1}$.
More generally, one could take $F(.)$ to be the 
cdf of a variable that follows some other distribution suitable for modeling count data, 
such as the negative binomial. Vectors of latent variables underlying counts  
$\uy^{(a)}_{i}=(y^*_{i1},\dots,y^*_{ip_{1}})^T$ are assumed to independently follow a 
$N_{p_1}(\mathbf{0},\uSigma_{i}^{(a)})$
distribution, where $\uSigma_{i}^{(a)}$ is restricted to be a correlation matrix since the variance parameters 
are non-identifiable by the data. The present model formulation allows for non-zero correlations among count 
outcomes \citep{vOphem99}. 

Concerning binomial responses, with relevant subscript in the range $k=p_{1}+1,\dots,p_1+p_{2}$, we let
$y_{ik} = t_2(y^*_{ik},\pi_{ik}) =\sum_{q=0}^{N_{ik}} q I[c_{ik,q-1} < y^*_{ik} < c_{ik,q}]$,
where $N_{ik}$ is the number of binomial trials, 
$c_{ik,-1} = -\infty$, and for $l \geq 1$,  
$c_{ik,l} = c_{l}(\pi_{ik}) = \Phi^{-1}\{G(l;N_{ik},\pi_{ik})\}.$
Here $G(.;N,\pi)$ is the cdf of a Binomial$(N,\pi)$ variable.
Note that, marginally, $Y_{ik} \sim \text{Binomial}(N_{ik},\pi_{ik})$.
Vectors of latent variables 
$\uy^{(b)}_{i}=(y^*_{i,p_{1}+1},\dots,y^*_{i,p_1+p_{2}})^T$ are assumed
to be independently distributed as $N_{p_2}(\mathbf{0},\uSigma_{i}^{(b)})$, where $\uSigma_{i}^{(b)}$, 
due to identifiability constraints, is a correlation matrix.

Lastly, for continuous responses $y_{ik}$, $k=p_1+p_{2}+1,\dots,p$, the corresponding latent variables 
are directly observed, $y_{ik} = y^*_{ik}$. The distributional assumption about vectors 
$\uy^{(c)}_{i} = (y^*_{i,p_1+p_{2}+1},\dots,y^*_{i,p})^T, i=1,\dots,n,$ is that they 
are independent $N_{p_3}(\ualpha_{i},\uSigma_{i}^{(c)})$ variates.

We let $\uy^{*}_{i} = \{(\uy^{(a)}_{i})^T,(\uy^{(b)}_{i})^T,(\uy^{(c)}_{i})^T\}^T$ 
denote the vector of latent variables underlying responses at location $i$.
It is assumed that the elements of $\uy^{*}_{i}$ jointly follow a multivariate normal distribution with 
mean parameter $\umu_{i}^{(y)}=(\mathbf{0},\mathbf{0},\ualpha_{i})^T$, and
block covariance matrix $\uSigma_{i}^{(y)}$ with diagonal blocks $\uSigma_{i}^{(a)}$, $\uSigma_{i}^{(b)}$, and
$\uSigma_{i}^{(c)}$ defined earlier, and 
with off diagonal blocks that represent covariances among latent variables
underlying different types of responses.   

Further, Poisson rates $\gamma_{ik}$, binomial probabilities $\pi_{ik}$, and continuous variable means 
$\alpha_{ik}$ are expressed in terms of risk factors $\ux_{ik}, i=1,\dots,n, k=1,\dots,p$, using canonical link functions \citep{glm}:
$\log(\gamma_{ik}) = \ux_{ik}^T \ubeta_{ik}$,  logit$(\pi_{ik}) = \ux_{ik}^T \ubeta_{ik}$, and $\alpha_{ik} = \ux_{ik}^T \ubeta_{ik}$.
In the sequel, we will use symbol $\ux_i$ to denote all risk factors that correspond to responses 
observed on area $i$ and symbol $\ubeta_i$ to denote the corresponding effects. Further, we will let
$\ubeta = \{\ubeta_{i}: i=1,\dots,n\}$.

We adjust for the effects of confounding variables $\uw_{i}$ by including them in the model in  
a similar way as the responses, but without modeling parameters of their distributions 
in terms of risk factors. Confounding variables can also be of mixed type. 
Specifically we assume that vector $\uw_{i}$ includes $q_1$ count, $q_2$ binomial, and $q_3$ 
continuous variables. Joint modeling is again facilitated by a latent variable representation. 
Hence, similar to $\uy^{*}_{i}$,
$\uw^{*}_{i}$ represents the vector of latent variables underlying confounding variables observed 
at location $i$,
and it is assumed to have a Gaussian distribution with mean $\umu_{i}^{(w)}$ and covariance 
matrix $\uSigma_{i}^{(w)}$.   

Jointly, $\uy^{*}_{i}$ and $\uw^{*}_{i}$ are assumed to follow a Gaussian distribution with mean 
$\umu_{i}^* = (\umu_{i}^{(y)},\umu_{i}^{(w)})$ and covariance matrix $\uSigma_{i}^*$ that has diagonal 
blocks $\uSigma_{i}^{(y)}$ and $\uSigma_{i}^{(w)}$, while its off diagonal block represents covariances
among the two sets of latent variables, 
cov$(\uy^{*}_{i},\uw^{*}_{i})$. We denote $\umu^{(w)} = \{\umu^{(w)}_{i}, i=1,\dots,n\}$
and $\uSigma^* = \{\uSigma_{i}^*, i=1,\dots,n,\}$.
Further, Poisson rates $\ugamma_i^{(w)}=(\gamma_{i1}^{(w)},\dots,\gamma_{iq_1}^{(w)})^T$
and binomial probabilities $\upi_i^{(w)}=(\pi_{i1}^{(w)},\dots,\pi_{iq_2}^{(w)})^T$ of confounding variables will collectively be denoted by
$\ugamma^{(w)}$ and $\upi^{(w)}$. 

With $\utheta_i=(\ubeta_i,\uSigma_i^*,\umu_i^{(w)},\ugamma_i^{(w)},\upi_i^{(w)})$ denoting the parameters of area $i$, 
the joint density of $(\uy_{i},\uw_{i})$ takes the form
\begin{eqnarray}\nonumber
 f(\uy_{i},\uw_{i}|\ux_{i};\utheta_i) = \int \dots \int N(\uy_{i}^{*},\uw_{i}^{*}|\umu_{i}^*,\uSigma_i^*) d\uy_{i}^{*} d\uw_{i}^{*},
\end{eqnarray} 
where the integral is with respect to latent variables underlying Poisson and binomial counts of
response and confounding variables, with integral limits that depend on $(\ubeta,\ugamma^{(w)},\upi^{(w)})$.

For all areas it is assumed that $(\uy_{i},\uw_{i})$ arises from a convolution density of the form
\begin{eqnarray}
 f_{i}(\uy_{i},\uw_{i}|\ux_{i}) = \int f(\uy_{i},\uw_{i}|\ux_{i};\utheta) dP_i(\utheta), \label{one}
\end{eqnarray} 
where $P_i(.)$ are location specific mixing distributions 
that are regarded as unknown and thus assigned a probit stick breaking process prior \citep{RD11}. 
Hence, they are represented as 
\begin{eqnarray}
 P_i(.)=\sum_{h=1}^\infty \pi_{hi} \delta_{\utheta_h}(.),\label{two}
\end{eqnarray}
where the atoms $\utheta_h=(\ubeta_h,\uSigma_h^*,\umu^{(w)}_h,\ugamma^{(w)}_h,\upi^{(w)}_h)$
are assumed to independently arise from the 
base distribution $G_0$ which consists of independent priors. More details on these priors are 
provided in Section \ref{mcmc}.

\subsection{Probit stick-breaking process priors}

Spatial dependence among measurements at nearby locations is induced by constructing the weights of the 
stick-breaking processes as probit transformations of latent variables that arise from Gaussian Markov 
random fields \citep{GMRFbook}. These random fields are multivariate normal distributions
defined on an undirected graph with areas represented by nodes and neighboring areas 
connected by an edge. Here areas are taken to be neighbors if they are geographically
contiguous. 

Mixture weights are obtained as
\begin{eqnarray}
 \pi_{hi} = \Phi(\eta_{hi}) \prod_{l<h} \{1-\Phi(\eta_{li})\},\nonumber
\end{eqnarray}
where $\eta_{hi} = \alpha + u_{hi}/\phi$, 
and the Gaussian Markov field realizations $\uu_h=( u_{h1},\dots,u_{hn})^T$ are obtained as 
independent draws from $N_n(\uzero,\uQ_{\lambda}^{-1})$, $h \geq 1$. 
The precision matrix is given by
$ \uQ_{\lambda} = \lambda \uA + \uI_n$,
where the adjacency matrix $\uA = \{a_{ii'}\}_{i,i'=1}^n$ is defined as follows: $a_{ii}=\nu_i$, 
the number of neighbors of area $i$, and for $i \neq i'$, $a_{ii'}=-1$ if locations $i$ and $i'$
are neighbors, and $a_{ii'}=0$ otherwise \citep{FG2002}. 
Thus, the probability density function (pdf) of $\uu_h$, $h \geq 1$, can be expressed as  
\begin{eqnarray} \label{grf}
p(\uu_h|\lambda) 
&=&c(\lambda)\exp\left\{-\frac{1}{2} \uu_h^T \uQ_{\lambda} \uu_h \right\} \nonumber\\
&=& c(\lambda)\exp\left[-\frac{1}{2}\left\{\lambda
\sum_{i' \sim i}\{u_{hi}-u_{hi'}\}^2+
\sum_{i=1}^n u_{hi}^2\right\}\right],
\end{eqnarray}
where $\sum_{i' \sim i}$ denotes the sum over all pairs of neighbors. 
The normalizing constant $c(\lambda)$ is given by 
$c(\lambda) = (2\pi)^{-n/2} \prod_{i=1}^n (\lambda e_i + 1)^{1/2}$, 
where $e_1,\dots,e_n,$ denote the eigenvalues of the adjacency matrix $\uA$. 

The non-negative parameter $\lambda$ determines the spatial correlation 
among the elements of $\uu_h$, with higher values of $\lambda$
implying higher correlations, whereas the limiting case of $\lambda=0$ implies
independence among the elements of $\uu_h$. The magnitude of $\lambda$ also  
determines the amount of shrinkage of the adjacency matrix $\uA$ towards 
the identity matrix, $\uI_n$. The effect of this shrinkage is to ensure that precision 
matrix $\uQ_{\lambda}$
is positive definite. 

This model formulation allows for the possibility that 
observations that correspond to nearby areas are more likely to have similar values for the component weights
than observations from areas that are far apart. Although parameter $\lambda$ clearly 
determines the correlations among the elements of the GMRFs, correlations among component 
weights depend on the combinations of values of the parameters that govern the GMRFs:
$(\alpha,\phi,\lambda)$. For instance a high value of $\lambda$ combined 
with a high value of $\phi$ implies smaller correlations among the component 
weights than the correlations implied by a high value of $\lambda$ combined with a small value of $\phi$. 

\section{Prior specification and MCMC sampler}\label{mcmc}

We develop a sampler that uses ideas from the work of \citet{RD11} and
implements the label switching moves suggested by \citet{PR08}. We focus on the case where there is one response of each type
and $q$ continuous confounders. Samplers for more general models can be constructed as a direct generalization of the 
presented sampler.

With $\uy_{i}=(y_{i1},y_{i2},y_{i3})^T$ denoting the vector of count, binomial and continuous responses,
$\uy^{*}_{i}=(y^{*}_{i1},y^{*}_{i2},y^{*}_{i3})^T$ denoting the corresponding latent variables,
and $\uw_{i}=(w_{i1},\dots,w_{iq})^T$ denoting the vector of confounders observed on location $i$, $i=1,\dots,n$,
the model is formulated as
\begin{eqnarray}\label{wis}
\uv_i \equiv ((\uy^{*}_{i})^T, \uw_{i}^T)^T |\{\umu_i^*,\uSigma^*_i\}  \sim N_s\left(
\umu_i^*=
\begin{array}{cc}
\left( 
\begin{array}{l}
\ualpha_i \\
\umu_{i}^{(w)} \\
\end{array}
\right),
 &
\uSigma^*_i=\left[ 
\begin{array}{ll}
\uSigma_i^{(y)} &  \uC_{i} \\
\uC_{i}^T & \uSigma_i^{(w)}\\
\end{array}
\right]
\end{array}\right),
\end{eqnarray}
where $s=3+q$,  E$(\uy^{*}_{i})=\ualpha_i$, E$(\uw_{i})=\umu_i^{(w)}$,
var$(\uy^{*}_{i})=\uSigma_i^{(y)}$, var$(\uw_{i})=\uSigma_i^{(w)}$, and cov$(\uy^{*}_{i},\uw_{i})=\uC_i$. 
Recall that the first two elements of $\ualpha_i$ are constrained to be zero and the third one is modeled as 
$\alpha_{i3} = \ux_{i3}^T \ubeta_{i3}$.
Hence, the mean $\umu_i^*$ can be expressed as $\umu_i^* = \uX^*_i \uxi_i$, where
$\uxi_i=(\ubeta_{i3}^T,(\umu_i^{(w)})^T)^T$, and $\uX_i^*$ is a design matrix
the first two rows of which include only zeros in order to satisfy the requirement of zero means: $E(y^*_{i1})=0,$
$E(y^*_{i2})=0$. Further, the first two diagonal elements of $\uSigma_i^*$ are constrained to be one. 
Lastly, Poisson rates $\gamma_{i}$ and Binomial probabilities $\pi_{i}$
are modeled as: $\log(\gamma_{i}) = \ux_{i1}^T \ubeta_{i1}$ and logit$(\pi_{i}) = \ux_{i2}^T \ubeta_{i2}$,
$i=1,\dots,n.$

The joint density of the data observed on the $i$th location $(\uy_{i},\uw_{i})$ takes the form
\begin{eqnarray}\label{five}
 f(\uy_{i},\uw_{i}|\ux_{i};\utheta_i) = 
 \int_{\Omega_{i2}} 
 \int_{\Omega_{i1}} N_s(\uv_i|\umu_{i}^*,\uSigma_i^*) d y_{i1}^* dy_{i2}^*,
\end{eqnarray} 
where $\Omega_{i1} = (c_{i,1,y_{i1}-1}, c_{i,1,y_{i1}})$, $\Omega_{i2} = (c_{i,2,y_{i2}-1}, c_{i,2,y_{i2}})$, 
and $\utheta=(\ubeta_{i1},\ubeta_{i2},\uxi_i,\uSigma^*_i)$ denotes model parameters. 

\subsection{Posterior sampling}

First note that from (\ref{one}), or its special case (\ref{five}), and (\ref{two}), the density of $(\uy_{i},\uw_{i})$ can be expressed 
as a countable mixture of densities, which we approximate by a truncated mixture 
\begin{eqnarray}
f_i(\uy_{i},\uw_{i}|\ux_{i}) = \sum_{h=1}^T \pi_{hi} f(\uy_{i},\uw_{i}|\ux_{i};\utheta_h). \label{slice1}
\end{eqnarray}

Introducing the usual allocation variables $\delta_{i}$, model (\ref{slice1}) can equivalently be
written as
\begin{eqnarray}
&& \uy_{i},\uw_{i} | \utheta, \delta_{i}=k_{i} \sim f(\uy_{i},\uw_{i}|\ux_{i};\utheta_{k_{i}}), \nonumber\\ 
&& P(\delta_{i}=k_{i}|\ueta) =\pi_{k_{i}i}, k_{i}=1,2,\dots. \nonumber
\end{eqnarray}
Further augmenting with latent variables underlying discrete responses 
$\uy_{i,1:2}^* = (y_{i1}^*,y_{i2}^*)^T$, we obtain the `complete data' likelihood
\begin{eqnarray}
\ell(\{\uy_{i},\uw_{i},\delta_{i}=k_{i},\uy_{i,1:2}^*:i=1,\dots,n\}) = \nonumber\\
\prod_{i=1}^n \left\{ I[y_{i1}^{*} \in \Omega_{i1}] I[y_{i2}^{*} \in \Omega_{i2}]
N_{s}(\uv_i|\uxi_{k_{i}},\uSigma^{*}_{k_{i}})\pi_{k_{i}i}\right\},  \nonumber
\end{eqnarray}
and the sampler updates from  
$\pi(\utheta,\udelta,\ueta,\alpha,\phi,\lambda,\uy^*|\uy,\uw) \propto
g_1(\uy|\uy^*,\udelta,\utheta) g_2(\uy^*,\uw|\udelta,\utheta)$ \\
$g_3(\udelta|\ueta) g_0(\utheta,\ueta,\alpha,\phi,\lambda) \propto$ 
\begin{eqnarray}
&& \prod_{i=1}^n \Big\{
I[c_{y_{i1}-1}(E_{i}\gamma_{k_{i}}) < y_{i1}^{*} < c_{y_{i}}(E_{i}\gamma_{k_{i}})]\nonumber\\
&& I[c_{y_{i2}-1}(\pi_{k_{i}}) < y_{i2}^{*} < c_{y_{i}}(\pi_{k_{i}})] 
N_{s}(\uv_i|\uxi_{k_{i}},\uSigma^{*}_{k_{i}})
\pi_{k_{i}i}\Big\} 
g_0(\utheta,\ueta,\alpha,\phi,\lambda), \nonumber
\end{eqnarray}
where $E_i$ denotes the expected number of counts in area $i$. Further details on the MCMC steps are provided in the Appendix. 

Prior specification $g_0(\utheta,\ueta,\alpha,\phi,\lambda)$ utilizes independent priors for parameters $\ubeta_{h1}$, $\ubeta_{h2}$, 
$\uxi_h$, and $\uSigma_h^*$, $h \geq 1$. We describe these in the following subsection. Priors for other parameters
can be found in the Appendix. 

\subsection{Specification of the base distribution and hyperparameters}

First, the priors for effects of the risk factors on the Poisson rates and binomial probabilities are specified as: 
$\ubeta_{hk} \sim N_{r_k}(\ubeta_{hk};\uzero,\tau^2\uI)$, where $r_k$ denotes the dimension, $k=1,2$. 
Similarly, the prior on $\uxi_h$ is taken to be 
$\uxi_{h} \sim N_{r_3+q}(\uxi_{h};\umu_{\xi},\uD_{\xi})$,
where $\umu_{\xi} = (\uzero^T,\uwbar^T)^T$. Here $\uwbar$
denotes the empirical mean of the confounding variables and $\uD_{\xi}$ is a diagonal matrix of   
$\tau^2$ (repeated $r_3$ times) followed by the empirical variances of the confounding variables. 
In our analyses we take $\tau^2=25$. 

We specify prior distributions on the restricted covariance matrices $\uSigma_{h}^{*}, h \geq 1,$
by incorporating additional variance parameters into the model that are non identifiable by the data \citep{xiao}
and separating identifiable from non identifiable parameters using the separation strategy of \citet{Barnard00}.
Specifically, we start by specifying Wishart$_s(\uE_h;\eta,\uEta)$ priors for unrestricted $s \times s$ 
covariance matrices $\uE_h, h \geq 1$: 
\begin{eqnarray}\label{wishart}
 p(\uE_h|\eta,\uEta) \propto |\uE_h|^{(\eta-s-1)/2} 
\text{etr}(-\uEta^{-1} \uE_h/2),
\end{eqnarray}
where etr$(.) = \exp(\text{tr}(.))$, and 
\begin{eqnarray}\nonumber
\uH =
\left[ 
\begin{array}{ll}
\uH_{11} & \uH_{12} \\
\uH_{12}^T & \uH_{22} \\
\end{array}
\right],
\end{eqnarray}
where $\uH_{11}$ is a $3 \times 3$ covariance matrix with its first two diagonal elements restricted to be
one, $\uH_{22}$ is a $q \times q$ unrestricted covariance matrix, and $\uH_{12}$ is a 
$3 \times q$ matrix of covariances.

We decompose $\uE_h=\uD_h^{1/2} \uSigma_h^{*} \uD_h^{1/2}$ into 
a diagonal matrix of two (non identifiable) variance parameters and $1+q$ ones 
(corresponding to identifiable variances), that is, $\uD_h = \text{Diag}(d^2_{h1},d^2_{h2},1,\dots,1)$, 
and a covariance matrix $\uSigma_h^{*}$ that has the required form.
The Jacobian that is associated with this transformation is 
$J(\uE_h \rightarrow \uD_h, \uSigma_{h}^{*}) = \prod_{j=1}^2 d_{hj}^{(s-1)} = |\uD_h|^{(s-1)/2}$,
and along with (\ref{wishart}) it implies a joint pdf for
$(\uD_h, \uSigma_{h}^{*})$: 
\begin{eqnarray}\label{wishart2}
 p(\uD_h, \uSigma_{h}^{*}|\eta,\uEta) \propto |\uE_h|^{(\eta-s-1)/2} 
\text{etr}(-\uEta^{-1} \uE_{h}/2) J(\uE_{h} \rightarrow \uD_h, \uSigma_{h}^{*}).
\end{eqnarray}
We take (\ref{wishart2}) to be the joint prior for $(\uD_h, \uSigma_{h}^{*})$.
Concerning posterior sampling, we will sample these two matrices together in a single
Metropolis-Hastings step, as was also done by \citet{xiao}.  

\section{Simulation studies}\label{simulation}

\subsection{First simulation study}

There are two main goals in the first of the two simulation studies that we present here. The first goal is to compare the proposed
model with the related models that were briefly described in the introductory section of the paper. 
These models will be described in more detail in the next few paragraphs. The second one is to appraise 
two of the aspects of the proposed model, namely the inclusion of the spatial structure 
and the continuous latent variable representation of the discrete variables, 
by comparing the model with 
special cases of it that do not take the spatial configuration into 
account and/or assume that discrete variables 
are conditionally independent.     

Both simulation studies are carried out on the spatial layout of the $n=94$ mainland French d\'epartments. 
In the first scenario that we present here a count response is assumed to be influenced by one continuous confounding 
variable and one continuous risk factor. Univariate
responses, risk factors and confounders will be denoted by $y_i, x_i,$ and $w_i,$
while expected counts will be denoted by $E_i$. The latter will be obtained as 
$E_i \stackrel{iid}{\sim} \text{Uniform}(10,20)$.

Given the above data specifications, the proposed model, which in the sequel we will denote by M$_1$, takes the 
form: $f_{i}(y_{i},w_{i}|x_{i}) = \sum_{h=1}^T \pi_{hi} f(y_{i},w_{i}|x_{i};\utheta_h)$.
Details on M$_1$ were provided in Section \ref{mcmc}. 
Here we note that the within component Poisson relative risks are expressed as 
$\gamma_{ih} = \exp(\beta_{0h} + \beta_{1h} x_i)$. 
We compare M$_1$ with the model proposed by \citet{Shahbaba} and \citet{Hannah}, denoted by M$_2$ and expressed as: 
$f_{i}(y_{i},w_{i}|x_{i}) = \sum_{h=1}^T \pi_{hi} g(y_{i}|x_{i},w_{i};\utheta_h') k(x_{i},w_{i}|\utheta_h')$.
Here $g(y_{i}|x_{i},w_{i};\utheta_h')$ denotes a Poisson pmf with relative risk 
$\gamma_{ih} = \exp(\beta_{0h} + \beta_{1h} x_i + \beta_{2h} w_i)$, while $k(x_{i},w_{i}|\utheta_h')$
denotes a bivariate Gaussian with unconstrained covariance matrix. 

We further consider two variations of M$_2$. The first one, which we will denote by M$_3$, imposes a 
diagonal covariance matrix in the multivariate Gaussian $k(|)$. 
The second one, denoted by M$_4$, imposes regression coefficients corresponding to confounding variables 
in the Poisson model $g(|)$ to take value zero. That is, M$_4$ sets $\beta_{2h} = 0$
for all $h$. As explained in the introduction, with these two constraints we attempt to mitigate the 
problems caused by high correlations and/or complex interactions, firstly by decomposing the 
overall dependence into clusters (M$_3$) and secondly by removing confounding variables
from the Poisson model and adjusting for their effects through the multivariate Gaussian (M$_4$).   

In addition, we consider a similar model to the one proposed by \citet{FG2002} and 
\citet{GR2002}, which takes the form 
$f_{i}(y_{i}|x_{i},w_{i}) = \sum_{h=1}^T \pi_{hi} f(y_{i}|x_{i},w_{i};\utheta_h^*)$.
This model is a countable mixture of Poissons, where 
the component specific relative risks are expressed as 
$\gamma_{ih} = \exp(\beta_{0h} + \beta_{1h} x_i + \beta_{2h} w_i)$.
We will denote this model by M$_5$.

Lastly, we consider two spatially varying coefficient models. At the observed level, both models are 
expressed as: 
\begin{eqnarray}
 Y_i &\sim& \text{Poisson}(E_i \lambda_i), \nonumber\\
 \log(\lambda_i) &=& \beta_{0i} + \beta_{1i} X_i + \beta_{2i} W_i, i=1,\dots,n. \nonumber
\end{eqnarray}
Let $\beta_{ki} = \beta_k + b_{ki}, k=0,1,2,$ and $\ub_i = (b_{0i},b_{1i},b_{2i})^T, i=1,\dots,n$. 
In the first specification, $\ub_i$ are modeled using an improper multivariate conditionally autoregressive (CAR) distribution:
\begin{eqnarray}
 \ub_i | \{\ub_j, j \neq i\}, \uOmega^{-1} \sim N_3(n_i^{-1}\sum_{j \sim i} \ub_j,n_i^{-1}\uOmega^{-1}).\nonumber
\end{eqnarray}
The second specification is a special case of the first one, where the common precision matrix $\uOmega$ is taken
to be diagonal, i.e. it is assumed that the coefficients of different covariates are independent. 
The two models will be denoted by M$_6$ and M$_{6A}$ respectively.

Model M$_1$ takes the spatial configuration into account and it also allows for non-zero within 
cluster correlation between continuous confounding and discrete response variables. We assess these two features of the 
model by comparing its performance with the performances of three models that are special cases of it, namely:
\begin{enumerate}
\item[M$_{1A}$:] a model that ignores possible spatial dependence by placing a degenerate at zero prior 
distribution on parameter $\lambda$, but that allows for non-zero within cluster correlation between 
confounding and response variables,
\item[M$_{1B}$:] a model that takes into account possible spatial dependence, but that assumes within 
cluster independence among confounding and response variables, that is, a model that describes 
component $h$ using the product density 
$f(y_i,w_i|x_i;\utheta_h)=\text{Poisson}(y_i|x_i;\gamma_h) N(w_i|\mu_h,\sigma^2_h)$, and  
\item[M$_{1C}$:] a model that ignores possible spatial dependence and that assumes within cluster 
independence among confounding and response variables. 
\end{enumerate}

We compare the models on the basis of their ability to recover the spatially varying 
risk factor effects. For this comparison we utilize the posterior mean squared error (MSE) that quantifies 
the discrepancy between true $\beta_{1i}$ and estimated $\hat \beta_{1i}$ risk factor effects:
MSE$(\beta_{1i}) = E\{(\beta_{1i}-\hat \beta_{1i})^2|\text{data}\}, i=1,\dots,n$.
As a one number summary that captures the performance of the models over the whole 
map, we calculate the root averaged mean squared error: RAMSE$(\ubeta_1)= (\sum_i \text{MSE}(\beta_{1i})/n)^{1/2}$.
Similarly, we calculate summaries over selected clusters of geographical areas. 

\begin{figure}
\begin{center}
\begin{tabular}{cc}
\includegraphics[width=0.30\textwidth]{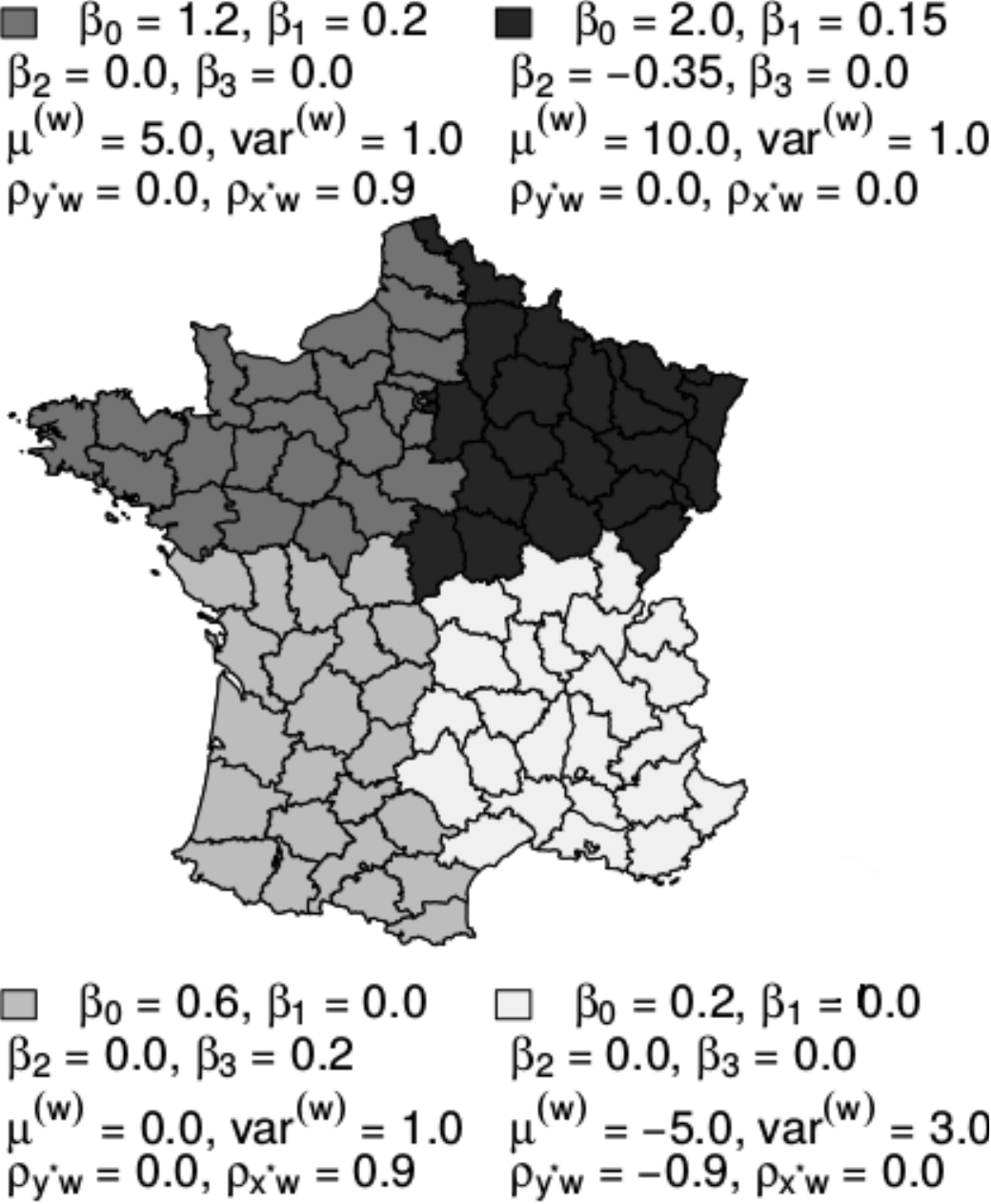} &
\includegraphics[width=0.40\textwidth]{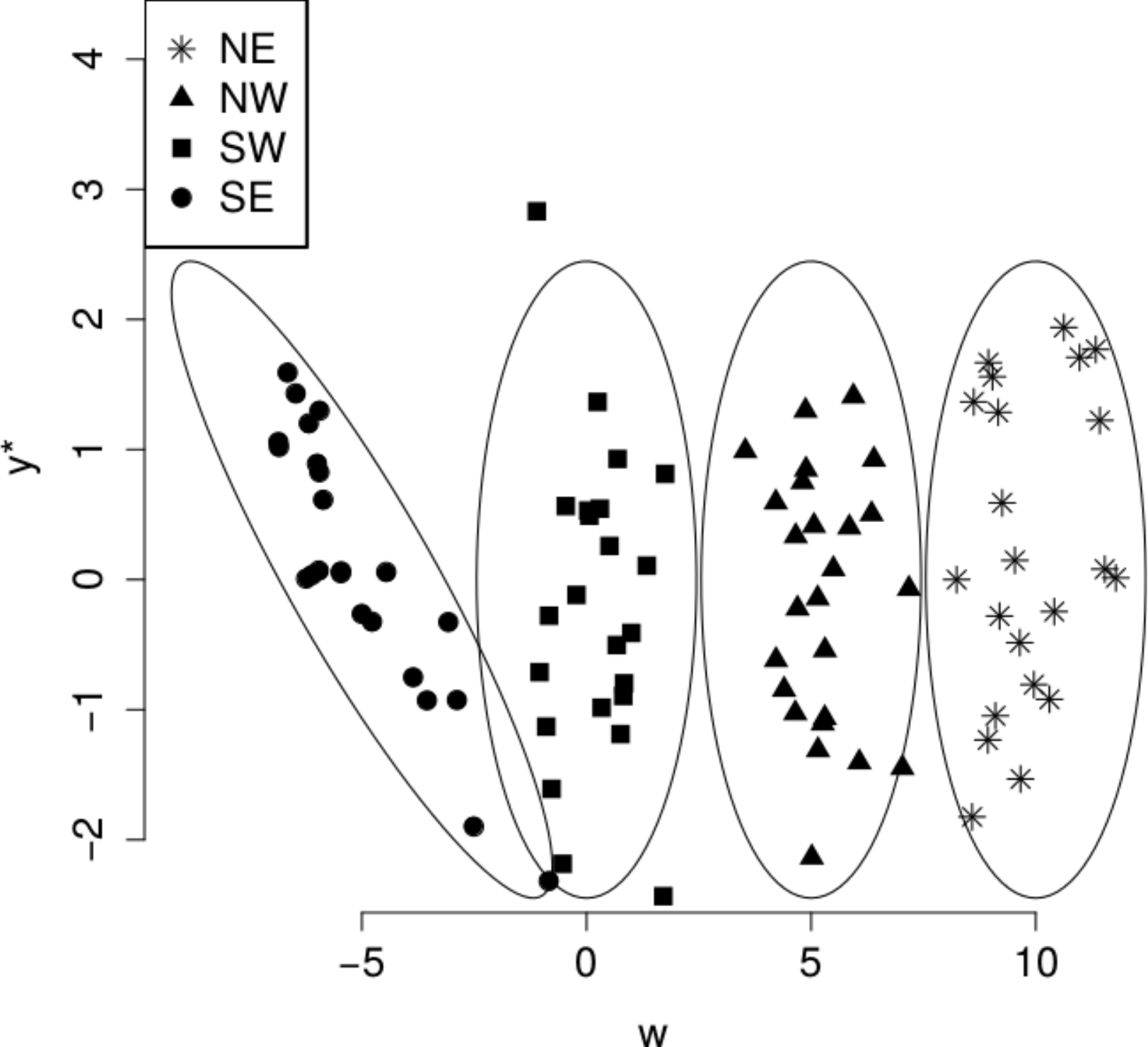}\\
(a) & (b)\\
\includegraphics[width=0.40\textwidth]{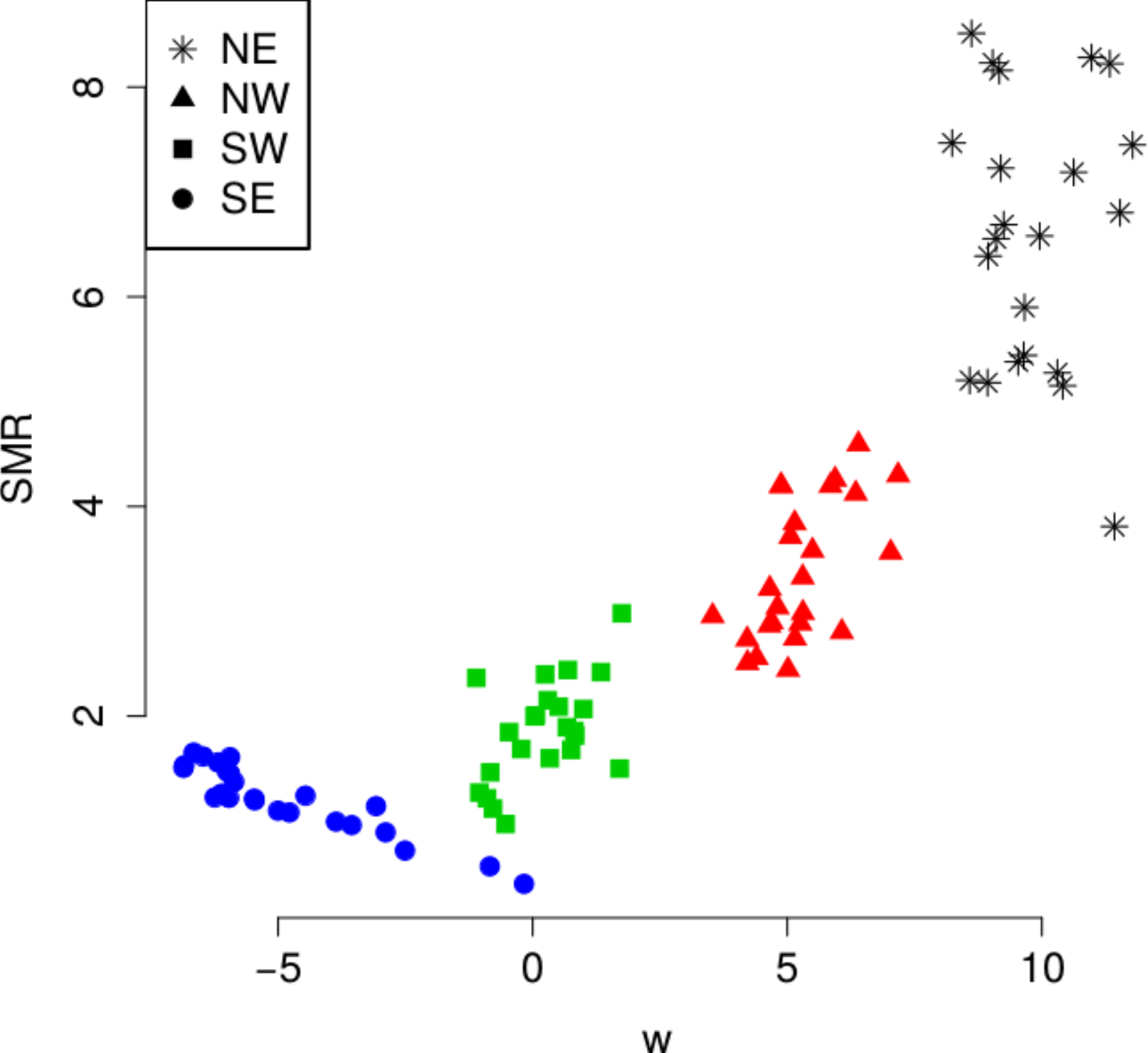} &
\includegraphics[width=0.40\textwidth]{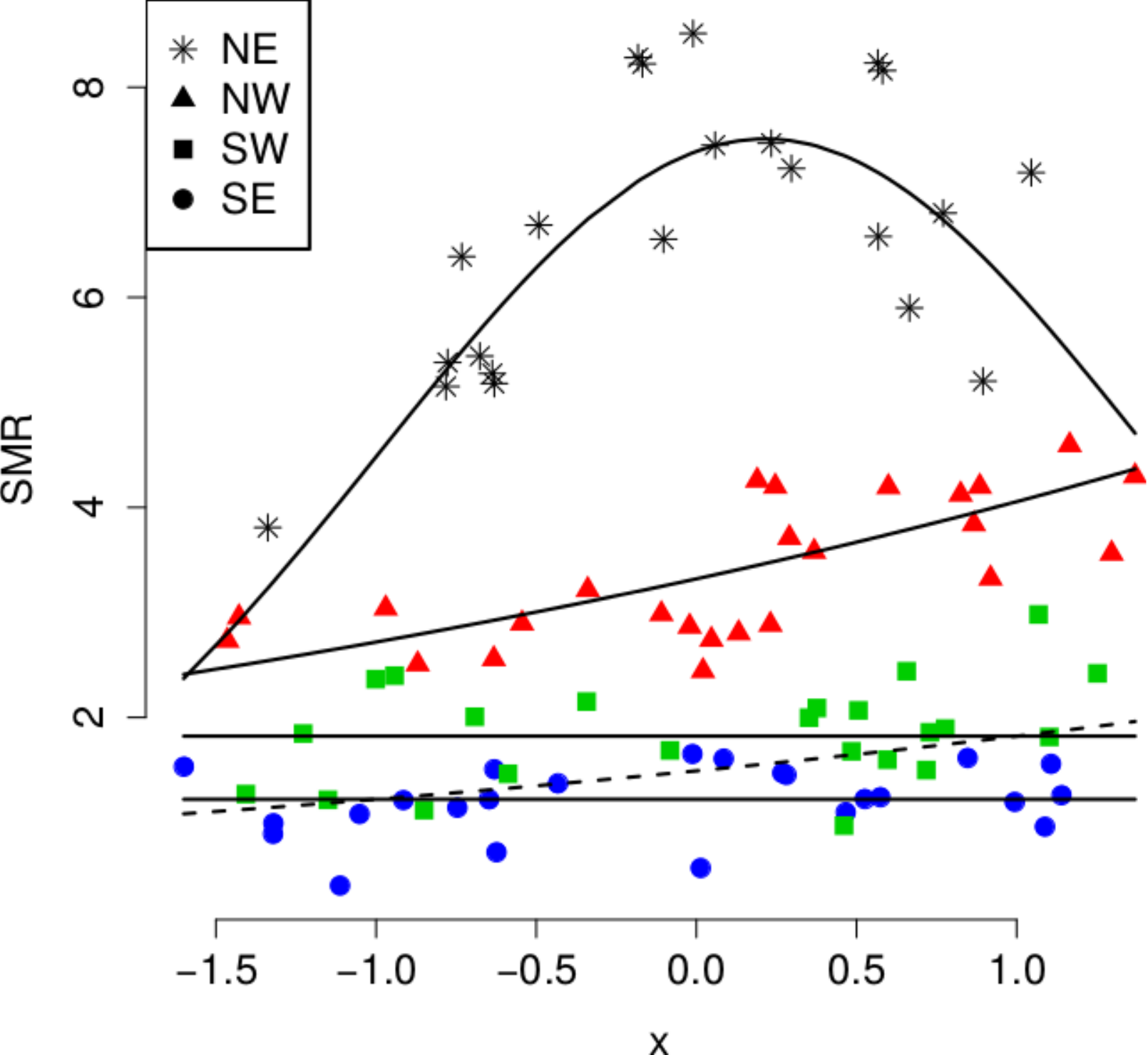}\\
(c) & (d)\\
\end{tabular}
\end{center}
\caption{First simulation study cluster structure and data. (a) Map of mainland French departments 
divided into four clusters and true cluster parameters. Parameters $\mu^{(w)}$ and 
var$^{(w)}$ are the mean and variance of the confounding variables. Parameter
$\rho_{y^*w}$ denotes the correlation between latent variables underlying counts and confounding variables,
while parameter $\rho_{x^*w}$ denotes the correlation between the variables that give rise to the risk
factors and the confounding variables. 
Parameters $\beta$ are the cluster specific intercepts and slopes. (b) Scatter plot of pairs of confounding 
variables $w$ and latent variables underlying counts $y^*$, along with $95\%$ ellipsoids of the 
respective Gaussians. (c) Scatter plot of SMRs against confounding variables. (d) 
Scatter plot of SMRs against risk factors, where the curves and lines display the true 
within cluster relationships between $x$ and SMR. The dashed line helps visualize the indirect positive relationship 
between $x$ and SMR within the SE cluster. Figures (b), (c) and (d) display data from one of thirty 
simulated datasets.}\label{true}
\end{figure}

In the current simulation study, the $n=94$ French departments were divided into four clusters. 
These are shown in Figure \ref{true} (a) along with the true model parameters.  
Thirty datasets (N=30) were generated by the following two stage process. At the first stage, continuous latent 
variables and risk factors, $y_i^*, x_i^*$, and directly observed confounding variables, $w_i,$ were obtained as 
realizations from a trivariate normal distribution. For instance, for the north-east (NE) cluster of areas, 
these were obtained from 
\begin{eqnarray}\nonumber
(y_{i}^{*},x_{i}^{*},w_{i})^T \stackrel{\text{iid}}{\sim} N_3\left(
\begin{array}{ll}
\left( 
\begin{array}{r}
0.0 \\
0.0 \\
10.0\\
\end{array}
\right),
 &
\left[ 
\begin{array}{ccc}
1.0 & 0.0 & 0.0\\
0.0 & 1.0 & 0.0\\
0.0 & 0.0 & 1.0\\
\end{array}
\right]
\end{array}\right).
\end{eqnarray} 
Parameters of the other three Gaussians are shown in Figure \ref{true} (a).
Note that, for all clusters we set E$(y_i^*)=0.0$, E$(x_i^*)=0.0$,
var$(y_i^*)=1.0$, var$(x_i^*)=1.0$, and cor$(y_i^*,x_i^*)=\rho_{y_i^*x_i^*}=0.0$,
and hence these are not shown on Figure \ref{true} (a). Figure \ref{true} (b) shows  
pairs $(w_i,y_i^*), i=1,\dots,94$, from one of the $30$ realized datasets, along with $95\%$ ellipsoids for the respective bivariate Gaussians.
 
At the second stage, risk factors, $x_i$, were obtained from $x_i^*$ as: $x_i = 3\Phi(x_i^*) - 3E\{\Phi(x_i^*)\}$,
so that $x_i \sim \text{Uniform}(-1.5,1.5)$. This two stage process allows the  
correlations cor$(y_i^*,x_i)$ and cor$(x_i,w_i)$ to be close to the desired ones, cor$(y_i^*,x_i^*)$ and cor$(x_i^*,w_i)$,
respectively. 
Furthermore, latent variables underlying counts were discretized using cut-points that respect the
desired relative risks and risk factor effects. Specifically,
Poisson counts were obtained as $y_i = q,$ where $q$ satisfies: 
$\Phi^{-1}\{F(q-1;E_i \exp(\beta_0 + \beta_1 x_i + \beta_2 x_i^2 + \beta_3 w_i))\} < y_i^* < 
\Phi^{-1}\{F(q;E_i\exp(\beta_0 + \beta_1 x_i + \beta_2 x_i^2 + \beta_3 w_i))\}$. 
Here $\Phi(.)$ and $F(.;.)$ denote the cdfs of the standard normal and Poisson distributions. 

The true model parameters, shown in Figure \ref{true} (a), have been chosen to provide enough separation 
among the four clusters in terms of their realized values of the confounding variables, while 
creating diverse within cluster relationships among the variables that will allow us to distinguish 
models in terms of their ability to cope with the challenges that these relationships bring. 
The NE cluster provides the challenge of the quadratic relationship between risk factor and response variable.  
The NW cluster creates the challenge of the high correlation between $x_i$
and $w_i$, with only $x_i$ having an effect on the response variable, $y_i$. Similar is the within SW
cluster challenge: there is high correlation between $x_i$ and $w_i$, but only $w_i$ has an effect on $y_i$.
Lastly, within the SE cluster, the response variable is only related to the confounding variable, not 
through the regression coefficient $\beta_3$, but through the high negative correlation between $w_i$
and $y_i^*$. To stress this relationship in the SE cluster, we have chosen the variance of  
$w_i$ to be higher than the corresponding variances within the other three clusters.  

For one of the $N=30$ generated datasets, Figure \ref{true} (c) presents a scatter plot of realized confounding variables against 
observed area relative risks. The latter are also known as standardized mortality ratios (SMRs), obtained as: SMR$_i = Y_i/E_i$. It is evident that there 
is enough separation among the four clusters in terms of $w_i$. Hence, models that cluster areas according to 
$w_i$, that is models M$_1$ - M$_4$, are expected to have an advantage over models that do not, that 
is models M$_5$ and M$_6$. We see that within the NE cluster, $y_i$ and $w_i$ are unrelated. Within the NW cluster, 
they are positively related. This is a result of the positive relationship between $y_i$ and $x_i$ and the 
positive relationship between $x_i$ and $w_i$. Within the SW cluster, the positive relationship between 
$y_i$ and $w_i$ is a result of the positive regression coefficient, $\beta_3 = 0.2$, while the negative 
relationship within the SE cluster is a result of the negative correlation between $y_i^*$ and $w_i$.

Similarly, Figure \ref{true} (d) is a scatter plot of the explanatory variable against the SMRs.
The Figure shows the quadratic relationship within the NE cluster, the linear
relationship within the NW cluster, and the lack of relationship within the SE cluster. 
Within the SW cluster, $x_i$ and $y_i$ are unrelated (indicated by the solid line) but 
the realized dataset shows a positive relationship due to the positive relationship between $y_i$ 
and $w_i$ and the positive relationship between $w_i$ and $x_i$ (indicated by the dashed line).    

\begin{figure}
\begin{center}
\begin{tabular}{cc}
\includegraphics[width=0.30\textwidth]{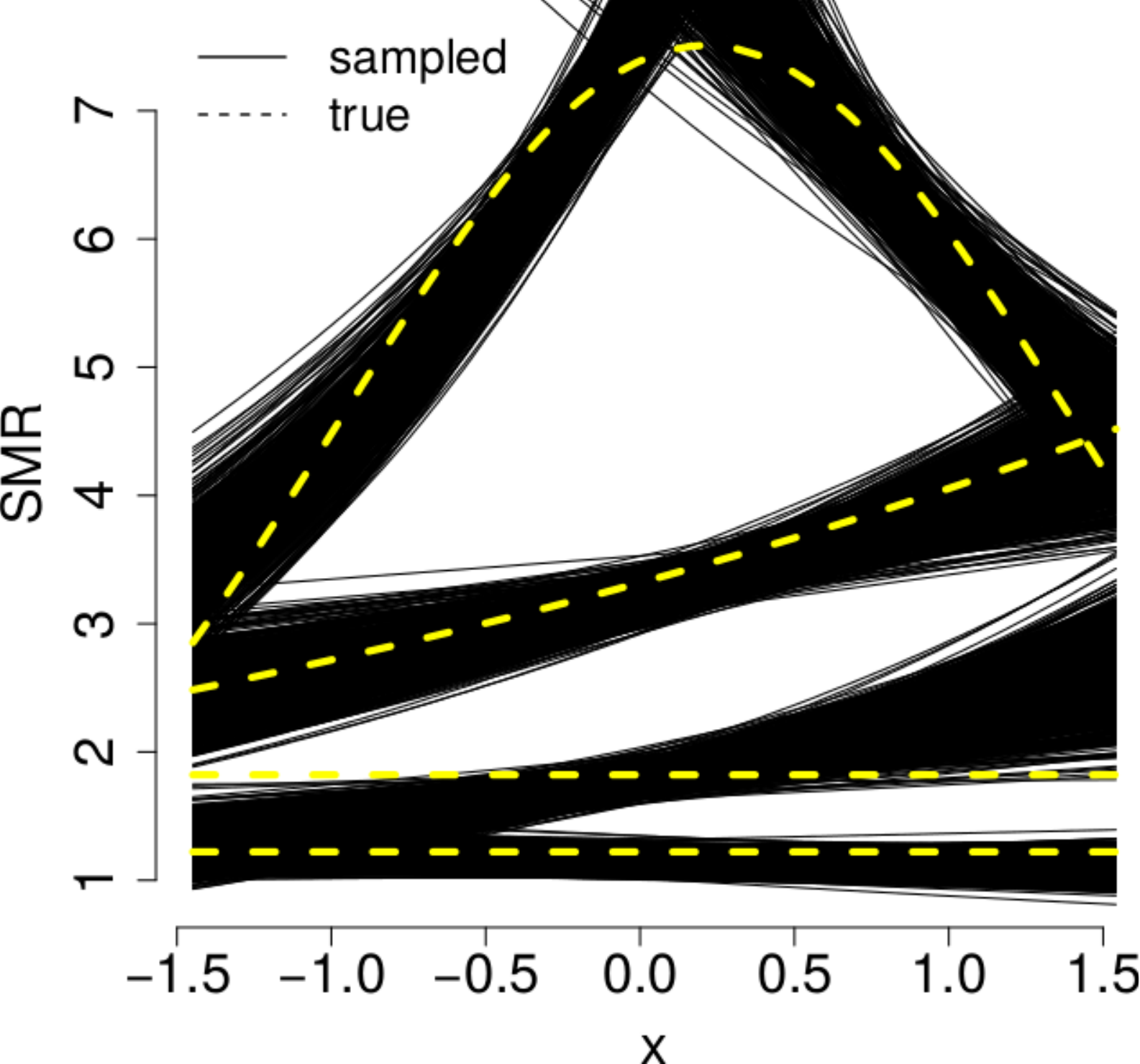} &
\includegraphics[width=0.30\textwidth]{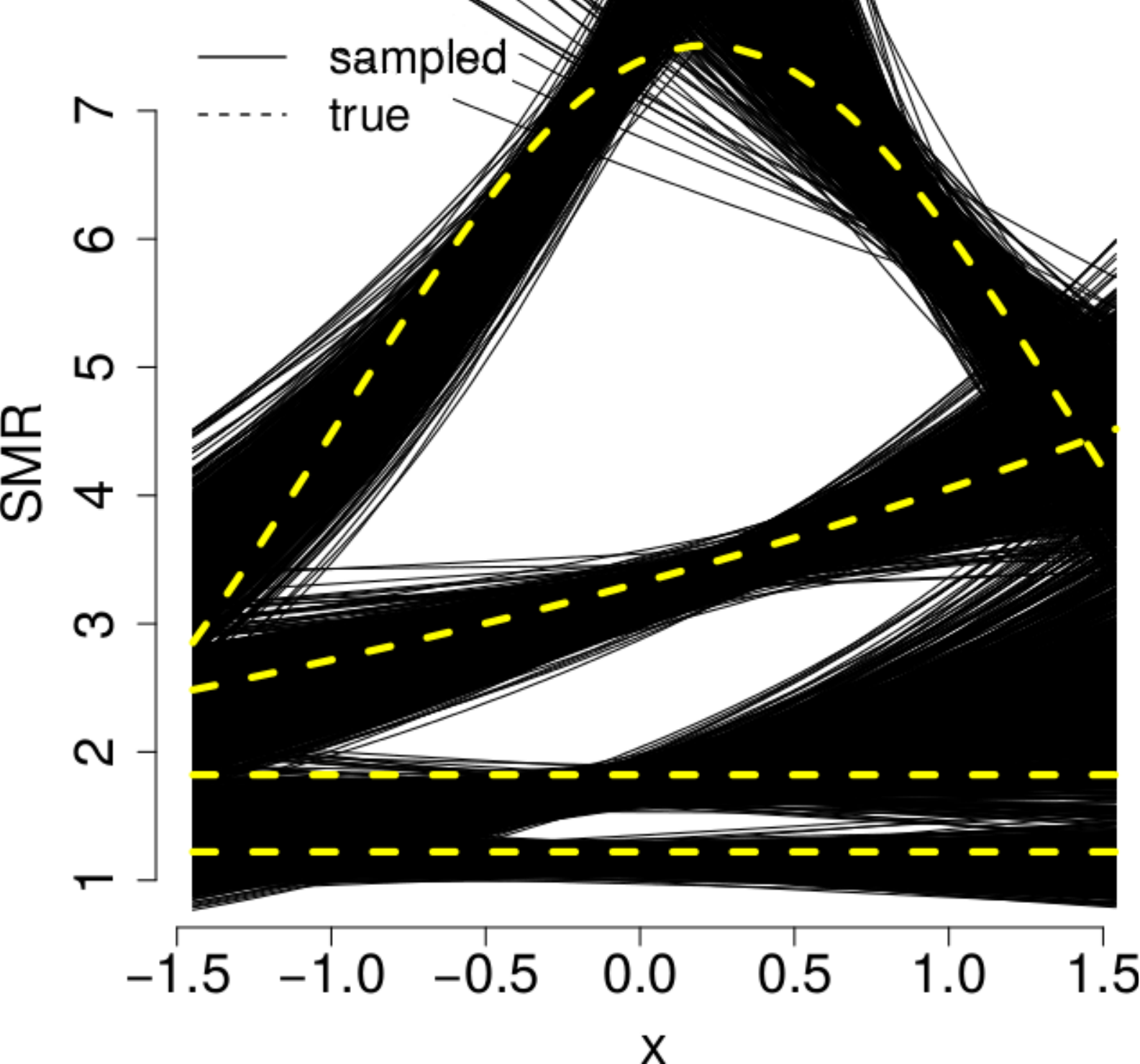} \\
M$_1$ & M$_2$\\
\end{tabular}
\end{center}
\caption{First simulation study results: within cluster model fits obtained at every 50th iteration 
of the samplers of models M$_1$ and M$_2$ for one of the thirty simulated datasets along with the true curves.}\label{it.spec.beta}
\end{figure}

\begin{figure}
\begin{center}
\begin{tabular}{cc}
\includegraphics[width=0.20\textwidth]{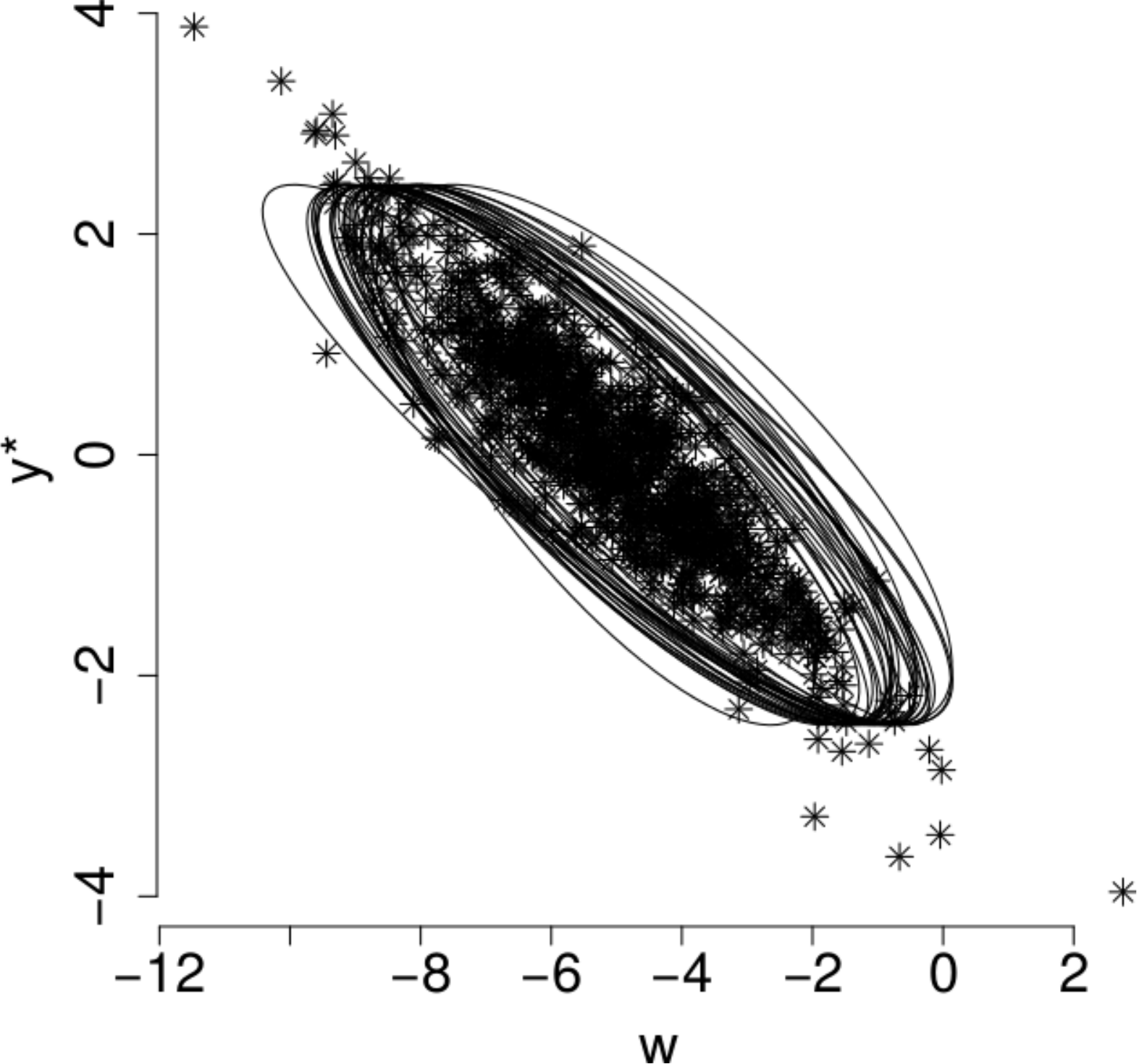} &
\includegraphics[width=0.20\textwidth]{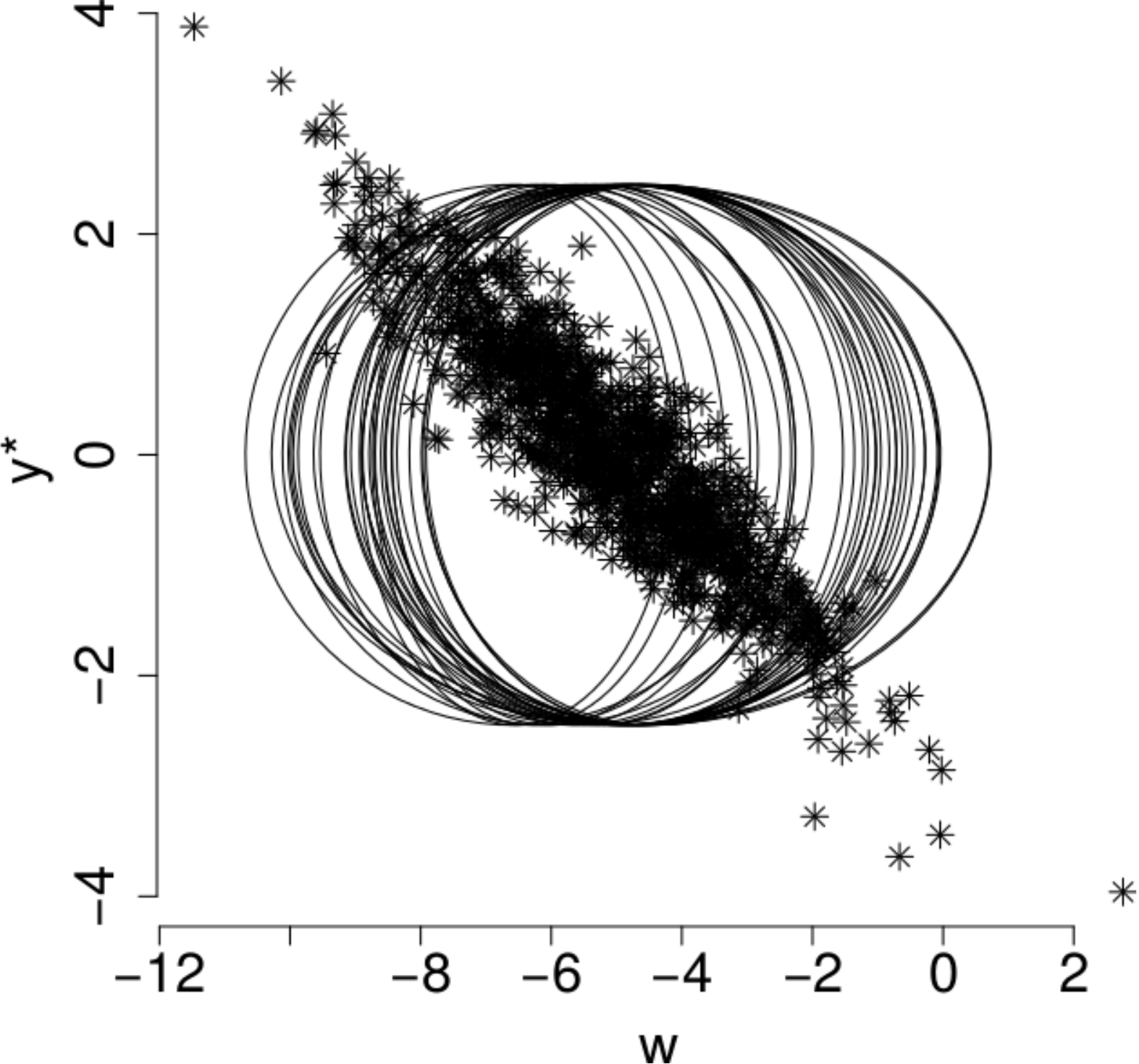} \\
(a) & (b)\\
\includegraphics[width=0.20\textwidth]{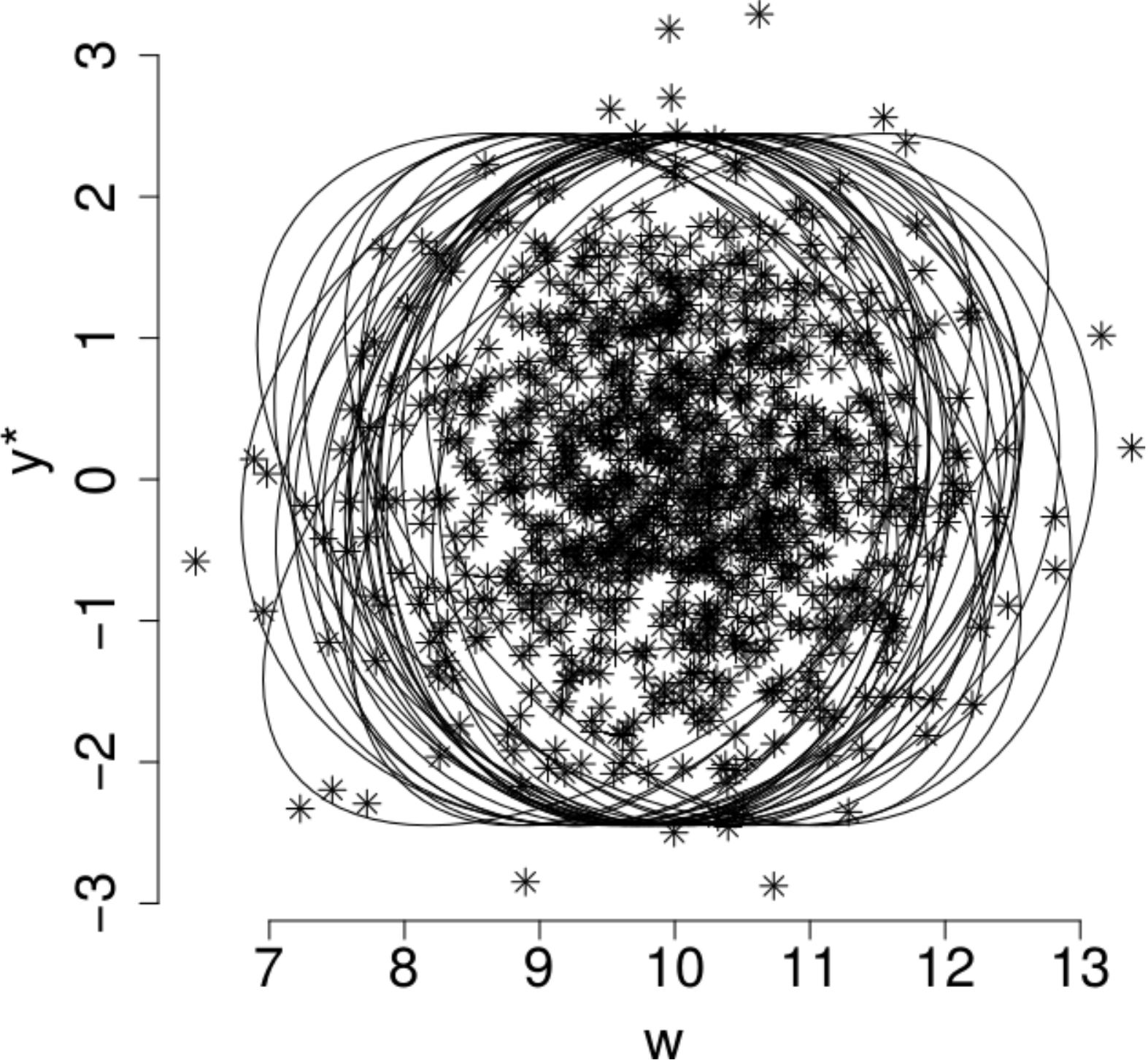} &
\includegraphics[width=0.20\textwidth]{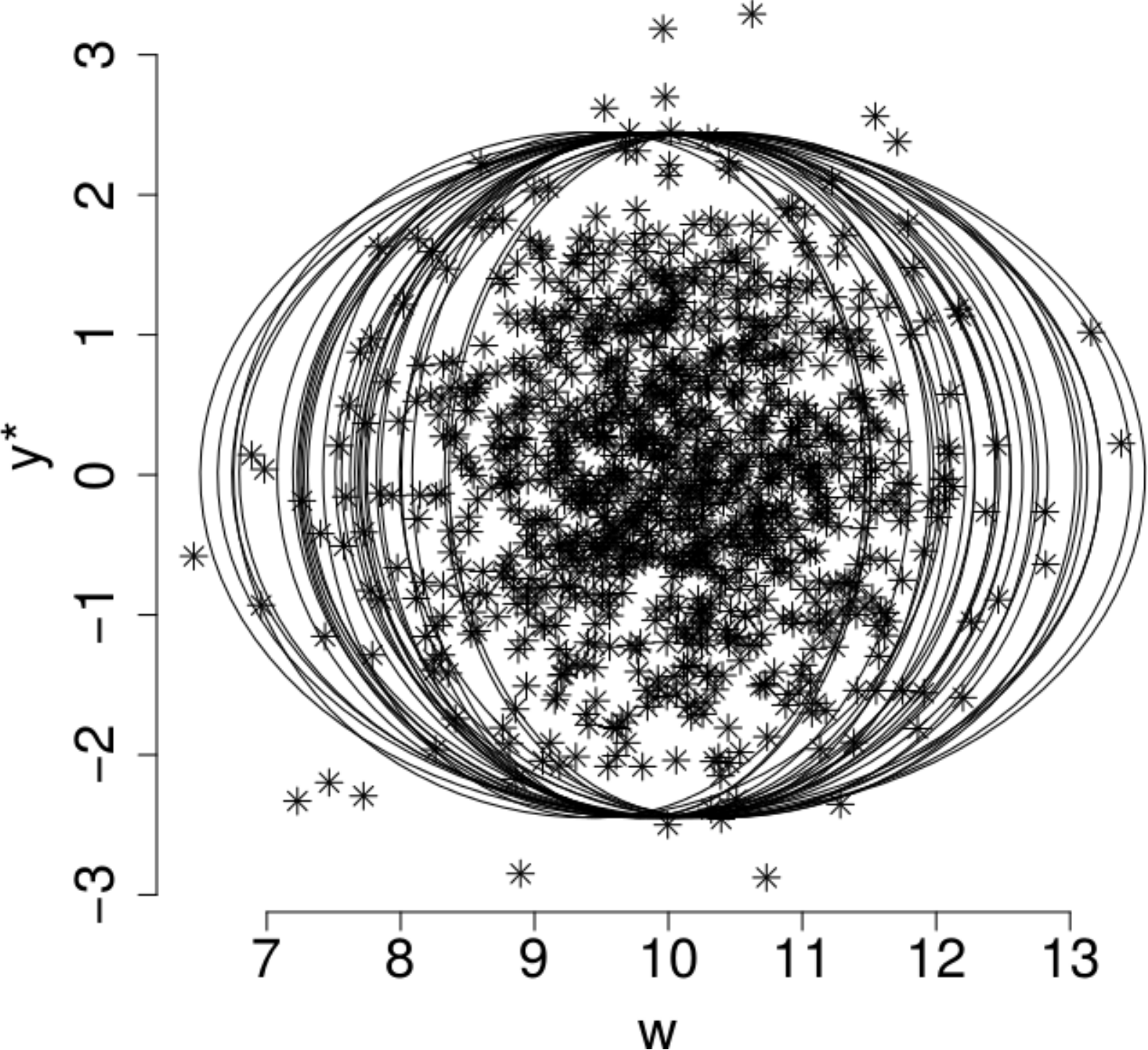} \\
(c) & (d)\\
\end{tabular}
\end{center}
\caption{First simulation study results: scatter plots of $(w,y^*)$ pairs obtained from the bivariate
Gaussians that describe the SE, (a) and (b), and NE, (c) and (d), clusters along with $95\%$ 
ellipsoids obtained from models M$_1$, (a) and (c), and M$_{1B}$, (b) and (d).}\label{latent}
\end{figure}

Results are obtained based on $50,000$ posterior samples, after a burn in period of $10,000$ samples, for each of the $N=30$ datasets, 
and are displayed in Figures \ref{it.spec.beta} and \ref{latent}, and Table \ref{table1.sim2}.
We first examine Figure \ref{it.spec.beta} and Table \ref{table1.sim2} that present summaries concerning the estimation 
of the spatially varying regression coefficients, and focus on comparing the performances of models M$_1$-M$_6$. 
Figure \ref{it.spec.beta} displays the within cluster curves that were obtained at every 50th iteration of  
the samplers of models M$_1$ and M$_2$, for one particular simulated dataset, along with the true curves. Note that, 
curves from model M$_3$ and M$_4$ are indistinguishable from those of models M$_2$ and M$_1$ respectively,
and hence not displayed. Further, model M$_5$ does not identify the clustering correctly and thus results from it are not displayed
either. It can be seen from Figure  \ref{it.spec.beta} that the quadratic relationship between explanatory variable and log SMR 
within the NE cluster is captured by splitting the cluster into $2$ sub-clusters: in one there is a positive linear relationship and 
in the other one a negative linear relationship. Within the NW cluster, that is characterized by a linear association between 
risk factor and log relative risk, and by high correlation between confounder and risk factor, models M$_1$ and M$_4$
that do not include the confounding variable in their linear predictors identify the true relationship with higher certainty than models M$_2$ 
and M$_3$. This is evident from both Figure \ref{it.spec.beta} and the first row of Table \ref{table1.sim2}. From the latter
we see that M$_1$ and M$_4$ have the smallest RAMSE$(\ubeta_1)$, while M$_3$ and M$_5$ have the highest, and M$_2$ and M$_{6A}$ 
have middle range RAMSEs. The RAMSE of M$_6$ is several times larger in every cluster than the RAMSE of every other model as it cannot cope with 
the high correlations between $x$ and $w$ that are present in some of the clusters. For this reason we exclude this 
model from further comparisons. 
Continuing with the SW cluster in which there is a linear relationship between confounding variable and log relative risk, and high 
correlation between confounding variable and risk factor, models M$_1$ and M$_4$ overestimate the regression coefficient $\beta_1$, 
that is they cannot
distinguish the causal effect from the effect that is due to the high correlation (Figure \ref{it.spec.beta}). 
Models M$_2$ and M$_3$, that include the confounding variable in their linear predictors, provide, due to 
multicollinearity, highly variable estimates of $\beta_1$ with posterior credible intervals that include the true value of the parameter. 
In terms of RAMSE$(\ubeta_1)$, M$_{6A}$ has the lowest while M$_3$ the highest. Lastly, within the SE cluster, in which there is no direct or indirect 
relationship between risk factor and risk, models M$_1$, M$_2$ and $M_{6A}$ do reasonably well in estimating $\beta_1$ (Figure \ref{it.spec.beta}),
with M$_1$ having the smallest RAMSE and M$_5$ the highest (Table \ref{table1.sim2}). 

Turning now to the comparison of M$_1$ with its three special cases, M$_{1A}$, M$_{1B}$, and M$_{1C}$,
we see from Table \ref{table1.sim2} that 
M$_{1A}$ that ignores spatial configuration does worse than M$_{1}$ in all geographical clusters. 
The obvious value of taking the geographical configuration into account in capturing the spatially varying coefficients 
is also illustrated by comparing models M$_{1B}$ and M$_{1C}$.
Comparing now model M$_{1B}$, that assumes within cluster independence,
with M$_{1}$, we see that M$_{1B}$ does better in the NW cluster where the assumption of local independence holds $(\rho_{y^*w}=\beta_3=0)$.  
However, M$_{1B}$ does worse than M$_1$ in the SW and SE clusters, where the assumption of local independence
does not hold. Comparison of the RAMSEs from models M$_{1A}$ and M$_{1C}$ further illustrates the points related to
local independence. 

\begin{table}
\begin{center}
\caption{First simulation study results: average RAMSE$(\ubeta_1)$ (over the $N=30$ simulated datasets) over the three geographical 
clusters where the relationship between log SMR and risk factor is linear.} \label{table1.sim2}
\begin{tabular}{lcccccccccc}
\hline
& M$_1$ & M$_2$ & M$_3$ & M$_4$ & M$_5$ & M$_6$ & M$_{6A}$ & M$_{1A}$ & M$_{1B}$ & M$_{1C}$\\
\hline
NW:   & 0.084 & 0.111 & 0.222 & 0.063 & 0.229 & 10.598 & 0.134 & 0.195 & 0.073 & 0.177\\
SW:   & 0.169 & 0.178 & 0.352 & 0.226 & 0.235 & 11.357 & 0.120 & 0.293 & 0.223 & 0.295\\
SE:   & 0.048 & 0.085 & 0.139 & 0.160 & 0.460 & 12.489 & 0.114 & 0.252 & 0.216 & 0.399\\
\hline
\end{tabular}
\end{center}
\end{table}

Lastly, Figure \ref{latent} shows the extra clustering flexibility gained by avoiding the assumption 
of local independence. Figures \ref{latent} (a) and (b) show realized pairs of $(w,y^*)$ from the bivariate normal 
density that describes the SE cluster of areas. Ignoring the location parameters, the bivariate 
normal has parameters var$(y^*)=1.0$, var$(w)=3.0$, and cor$(y^*,w)=-0.9$. 
Figure \ref{latent} (a) shows how model M$_{1}$ deals with this negative dependence. The Figure
displays, along with realized pairs, $95\%$ ellipsoids that were obtained in the simulation study from model M$_{1}$. 
Ignoring the dependence parameter, i.e. setting cor$(y^*,w)=0.0$, as in M$_{1B}$, results in a considerably worse fit, as
illustrated in Figure \ref{latent} (b). 

It is of course advantageous to assume local independence
when the variables are locally independent, as they are in the NE cluster.  
Adding, however, an extra parameter in the model to capture possible dependence, does not result in great loss. 
This is illustrated in Figures \ref{latent} (c) and (d) that display pairs of $(w,y^*)$ realizations from 
a bivariate normal with var$(y^*)=\text{var}(w)=1.0$ and cor$(y^*,w)=0.0$, along with $95\%$ 
ellipsoids obtained from models M$_1$ and M$_{1B}$ respectively. 

\subsection{Second simulation study}

The purpose of the second simulation study is to evaluate the non-parametric structure when the true
data generating mechanism is a generalized linear mixed effects model with random intercepts 
obtained as realizations form a GMRF, but with no further covariates or confounders. 

More comprehensive comparisons between CAR type models and mixtures of Poisson pmfs can be found in 
\citet{FG2002}, \citet{GR2002} and \citet{BRT2005}.  
These have concluded that $1.$ even when the true underling model is a CAR type model, mixture models
perform very competitively, and $2.$ when there are discontinuities in the risk surface, mixture models 
outperform CAR models as the latter lack a mechanism for dealing with gaps and hence they 
oversmooth the risk surface. Below we describe in more detail our simulation study. 

The synthetic datasets are obtained by a two stage process. At the first stage a GMRF $\uu$ is obtained from the 
proper pdf $p(\uu|\lambda)$ given in (\ref{grf}). At the second stage, count responses are generated as $Y_i \sim \text{Poisson}(E_i\exp(\eta_i))$,
where $\eta_i = u_i/\phi,$ $E_i \stackrel{iid}{\sim} \text{Uniform}(10,20)$, $i=1,\dots,n$.
We have selected four different values for $\lambda$ and $1/\phi$, which are shown
in Table \ref{sim2}. For each combination of the two parameters, we generated $N=20$ datasets
and fitted nonparametric and CAR models. 

The nonparametric model takes the form $f_i(y_i) = \sum_h \pi_{hi} g(y_i|\theta_h)$, where $g(.|\theta)$ denotes 
a Poisson pmf with relative risk $\theta$. This model is a special case of M$_{5}$ that has no covariates and it is reminiscent 
of the model proposed by \citet{FG2002}. 
The CAR model is expressed as $Y_i \sim \text{Poisson}(E_i\exp(\theta_i))$, 
where $\theta_i \sim N(n_i^{-1} \sum_{j \sim i} \theta_j, n_i^{-1}\tau^2), i=1,\dots,n$,
reminiscent of the model by \citet{BYM}. 

We summarized the performances of the two models calculating the 
RAMSE$=((Nn)^{-1}\sum_{k=1}^N\sum_{i=1}^n (\eta_i -\hat \eta_i)^2)^{1/2}$.

\begin{table}
\begin{center}
\caption{Second simulation study results: the first (second) entry in each cell is the RAMSE obtained from the 
nonparametric (CAR) model.} \label{sim2}
\begin{tabular}{cc|cccc|}
 & \multicolumn{1}{c}{} & \multicolumn{4}{c}{$1/\phi$}\\
 & \multicolumn{1}{c}{}          & 1 & $2^{1/2}$ & 2 & \multicolumn{1}{c}{$2^{3/2}$}\\ 
 \cline{3-6}
                            & 1  & 0.1085$|$0.0477 & 0.1482$|$0.0609 & 0.1680$|$0.0797 & 0.2702$|$0.1244\\ 
\multirow{2}{*}{$\lambda:$} & 5  & 0.0663$|$0.0297 & 0.0988$|$0.0434 & 0.1161$|$0.0529 & 0.1341$|$0.0594\\ 
                            & 10 & 0.0417$|$0.0202 & 0.0686$|$0.0316 & 0.1017$|$0.0404 & 0.0964$|$0.0385\\ 
                            & 20 & 0.0386$|$0.0139 & 0.0469$|$0.0217 & 0.0694$|$0.0295 & 0.0925$|$0.0404\\ 
\cline{3-6}
\end{tabular}
\end{center}
\end{table}

Results are displayed in Table \ref{sim2}, where the first (second) entry in each cell is the RAMSE obtained from the 
nonparametric (CAR) model.
RAMSEs obtained from the nonparametric model are, on average, 2.2 times larger than those obtained from the
CAR model, with small variation around this number. Lastly, it is interesting to observe that RAMSEs increase with increasing 
variance $1/\phi$ and decrease with increasing spatial association parameter $\lambda$. 

\section{An examination of the association between birth outcomes and exposure to ambient air pollution}\label{application}

We apply the proposed model to study the association between two birth outcomes and exposure to ambient air pollution.
The birth outcomes that we consider are preterm birth and birth weight. Both of these serve as proxy measures of the 
degree of biological maturity of the fetus for supporting extrauterine life. 
As a measure of air pollution we consider the total suspended particulate matter (PM) equal to or less than 10 
micrometers ($\mu$m) in diameter (PM$_{10}$).

Preterm birth is defined as delivery before 37 completed weeks of gestation (birth occurring at 
least four weeks before the estimated date of delivery). 
Determining, however, when natural conception takes place, and hence gestational age at birth, has been difficult. 
For this reason, birth weight was originally used as a proxy measure for maturity.
The main issue, however, with birth weight as a proxy of immaturity is that 
it may misclassify many infants, for instance those who have small/large weight for their gestational 
age. Hence, gestational age is considered as a better surrogate of maturity and it is preferred over
birth weight, whenever it is available (see e.g. \citet{BB}). Here, the response variables that we consider are 
the dichotomous $Y_1:$ gestational age at birth $\leq 37$ weeks, and the continuous $Y_2:$ birth weight. 

Several epidemiological studies have examined the relationship between environmental air pollution exposures and 
preterm birth and birth weight, 
with, however, unclear results. For instance, a recent systematic review and meta-analysis \citep{stieb} reports that the majority of the 
studies reviewed, found that increased air pollution was associated with reduced birth weight. However, the authors also reported 
evidence of publication bias. Further, the same authors reported that the estimated effects on preterm birth were mixed. Inconsistent results 
have also been reported elsewhere, see e.g. \citet{BB} and references therein. The majority of these studies considered   
birth weight as the response variable due to the difficulties with gestational age mentioned above. For instance in \citet{stieb}
there are 62 (8) studies that consider weight (gestational age) as the response. Here, we add to the literature a study 
that considers both responses simultaneously, and can thus shed light on how the air pollution effects on the two responses compare.     

There are several factors that can contribute to a premature birth and for which we adjust our analysis. 
Cigarette smoking has been associated with adverse pregnancy outcomes by a number of studies, although
reported results have not been entirely consistent (see e.g. \citet{BB}, pages 91-92). 
As smoking rates per area are not available in our study, we adjust for the effects of smoking by including in the model area level lung cancer occurrence counts. 
This is the first confounding variable that we include in the model, denoted by $W_1$, and it serves as proxy to smoking 
rates \citep{BestH}. 

In addition, several studies have documented significant associations between area-level characteristics and birth outcomes, see e.g. 
\citet{Elo} and \citet[pages 137-147]{BB} and references therein. Area-level characteristics such as crime rates and socioeconomic deprivation 
can influence health outcomes 
through pathways such as exposure to acute or chronic stress and availability of social support and goods and services. 
We account for area level characteristic by including in the model (sub)-domains of the Index of Multiple Deprivation 2010 (IMD) \citep{imd}. 
Specifically, we include the domains of `Income' deprivation, `Crime rates', `Distance to local services' 
(services such as general practice surgery and stores) and `Housing quality'. 
For all domains higher scores indicate relatively less advantaged areas. However, only `Income' deprivation scores are expressed in meaningful units. 
These represent proportions of income deprived people in the areas. The construction of 
all other domain scores, including the overall IMD score, involves an exponential transformation that results in deprivation scores that are 
difficult to interpret. We overcome this difficulty by ranking the domain scores and dividing the ranks by the total number of areas. 
These new scores are more meaningful: the score of a given area represents the proportion of areas that are less deprived than that area. 

Furthermore, there is evidence of significant differences in birth outcomes among different ethnic groups \citep{BB}. 
Hence, we adjust our analysis for area-wise ethnic distributions, expressed as percentages of people 
whose ethnic background can be described as White, Asian, or Other, denoted by $p_\text{w}, 
p_\text{a},$ and $p_\text{o}$. We include two of these percentages in the model
after applying a `logit' transformation: $p_\text{a}^* = \log\{p_\text{a}/p_\text{w}\}$,   
$p_\text{o}^* = \log\{p_\text{o}/p_\text{w}\}$. The purpose of these transformations is
to create variables that have the real line as their support so that they can be modeled
by a mixture of multivariate normal densities.

Lastly, as it is well known that maternal age can have important effects on birth outcomes (see e.g. \citep[pages 44-47]{BB})
we adjust our analysis for the area-wise mean maternal age. 

By utilizing the proposed model we can examine the effect of ambient air pollution on the two birth outcomes of interest 
while automatically adjusting via the clustering aspect of the model for the possibly nonlinear effects of the other risk factors 
and their interactions. The latter can be important in this application as nonlinear and interaction effects 
among the aforementioned risk factors have been described in the literature. 
For instance, \citet{GRA} found that racial/ethnic differences in birth weights become more pronounced as 
pregnancies approach term. In addition, nonlinear effects of maternal age on the risk of preterm birth have been described 
\citet[pages 125-127]{BB}: there is 
higher risk associated with young maternal ages and ages over 35. Furthermore, the effect of maternal age on preterm birth
varies among racial/ethnic groups. For instance, the risk of preterm birth starts to increase at a later age for 
whites than for blacks, and this increase is slower for whites. 

We examine whether maternal exposure to PM$_{10}$ increases the risk of adverse birth outcomes 
in a small area study involving the $n=628$ Output Areas (OA) of Greater London, $2008$.  
The two response variables that we consider are $Y_{i1}$ the number of preterm births in area $i$, 
and $Y_{i2}$ the average birth weight in area $i$. Note that, as multiple gestations is one of the strongest risk 
factors for premature birth, we confine our analysis to singleton births. Given the total number of singleton 
births per area, $N_i$, variable $Y_{i1}$ is modeled as $Y_{i1} \sim$Binomial$(\pi_i,N_i)$, where logit$(\pi_i) = \ux_{i1}^T \ubeta_{i1} = 
\beta_{i,01} + \beta_{i,11} \text{PM}_{10,i}$, where $PM_{10,i}$ is the estimated annual average exposure to PM$_{10}$
in area $i$. Variable $Y_{i2}$ is modeled as $Y_{i2} \sim N(\alpha_i,\sigma_{i2}^2)$,
where $\alpha_i = \ux_{i2}^T \ubeta_{i2} = \beta_{i,02} + \beta_{i,12} \text{PM}_{10,i} + \beta_{i,22} \text{O}_{i}$, with $O_i=Y_{i1}/N_i$ denoting the 
observed proportion of preterm births in area $i$. Hence, the model for the latent and observed continuous response variables $(y^{*}_{i1}, y_{i2})$ for area $i$ takes the form
\begin{eqnarray}\nonumber
(y^{*}_{i1}, y_{i2})^T |\{\ubeta_i,\uSigma^*_i\}  \sim N_2\left(
\begin{array}{cc}
\left[ 
\begin{array}{l}
\beta_{i,01} + \beta_{i,11} \text{PM}_{10,i} \\
\beta_{i,02} + \beta_{i,12} \text{PM}_{10,i} + \beta_{i,22} \text{O}_{i} \\
\end{array}
\right],
 &
\left[ 
\begin{array}{ll}
1.0 &  \sigma_{12} \\
\sigma_{21} & \sigma_{22}\\
\end{array}
\right]
\end{array}\right).
\end{eqnarray}
The inclusion of the proportion of preterm births $O_i$ as a covariate for birth weight $Y_{i2}$ defines a recursive model. 
Such models are extensively used in the econometric literature to adjusted for unobserved confounding, see e.g. \citet{Heckman} and 
\citet{Goodmanetal} for an application in biostatistics. 

The number of lung cancer occurrences, the first confounding variable $W_{i1}$, and age-sex distribution for each OA 
were available in 5-year age bands. The expected number of lung cancer occurrences, $E_i, i=1,\dots,n$, were calculated based on the age-sex distributions,
thereby adjusting for these two important risk factors. Counts where modeled as $W_{i1} \sim \text{Poisson}(E_i\gamma_i)$,
where $\gamma_{i} = \exp(\beta_{i,03})$. Additional confounders included in the model are the four IMD (sub)-domains, $W_{i2},\dots,W_{i5}$,
two variables describing the ethnic distribution, $W_{i6}$ and $W_{i7}$, and maternal age $W_{i8}$.

\begin{figure}
\begin{center}
\includegraphics[width=0.3\textwidth]{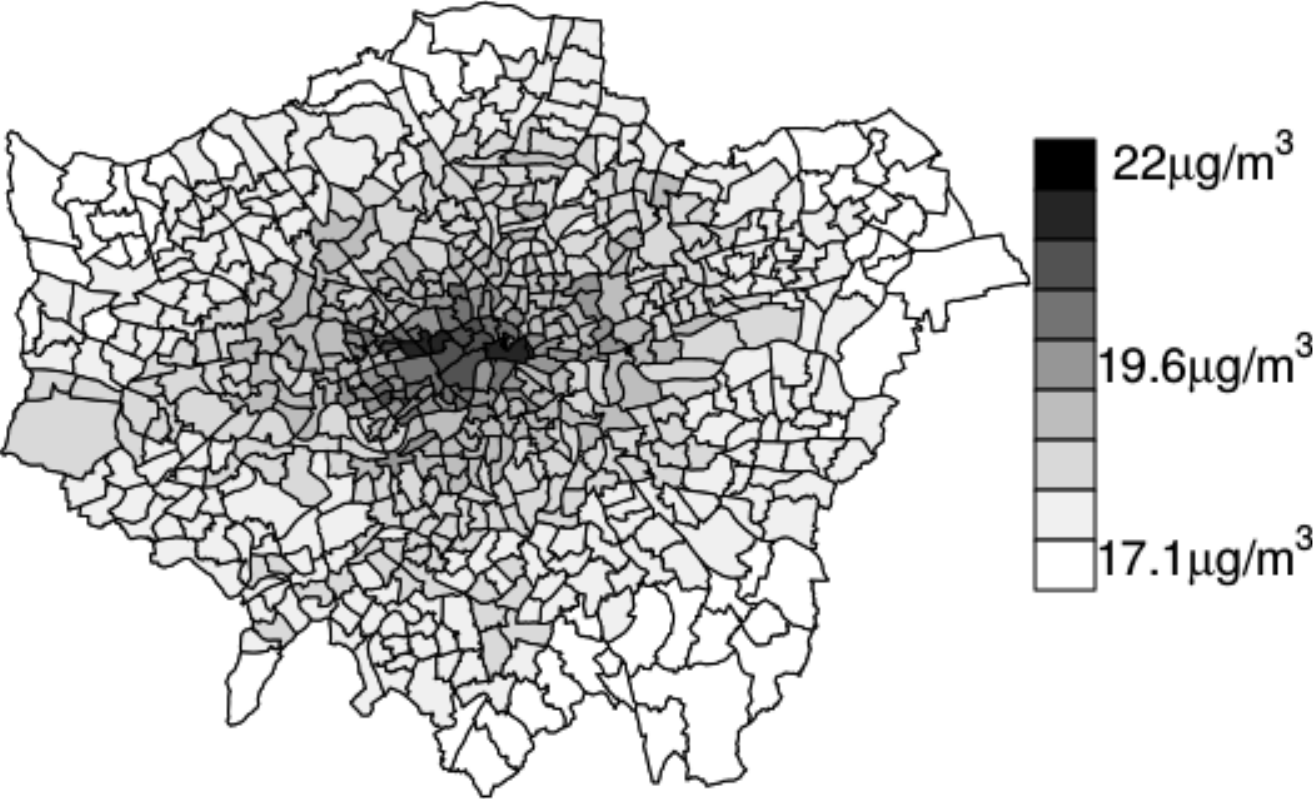} 
\end{center}
\vspace{-0.2cm}
\caption{Estimated average annual exposure to PM$_{10}$.}\label{pm10}
\end{figure}

\begin{figure}
\begin{center}
\begin{tabular}{cc}
\includegraphics[width=0.3\textwidth]{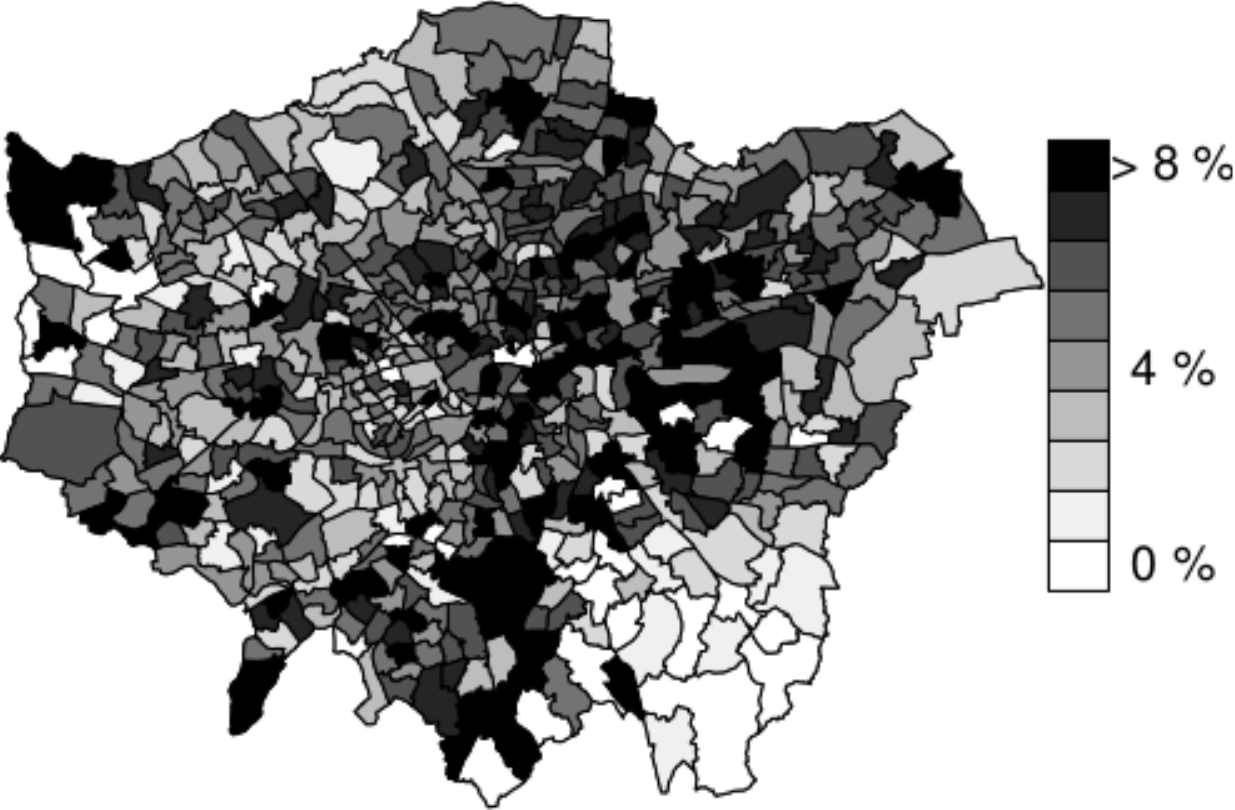} &
\includegraphics[width=0.3\textwidth]{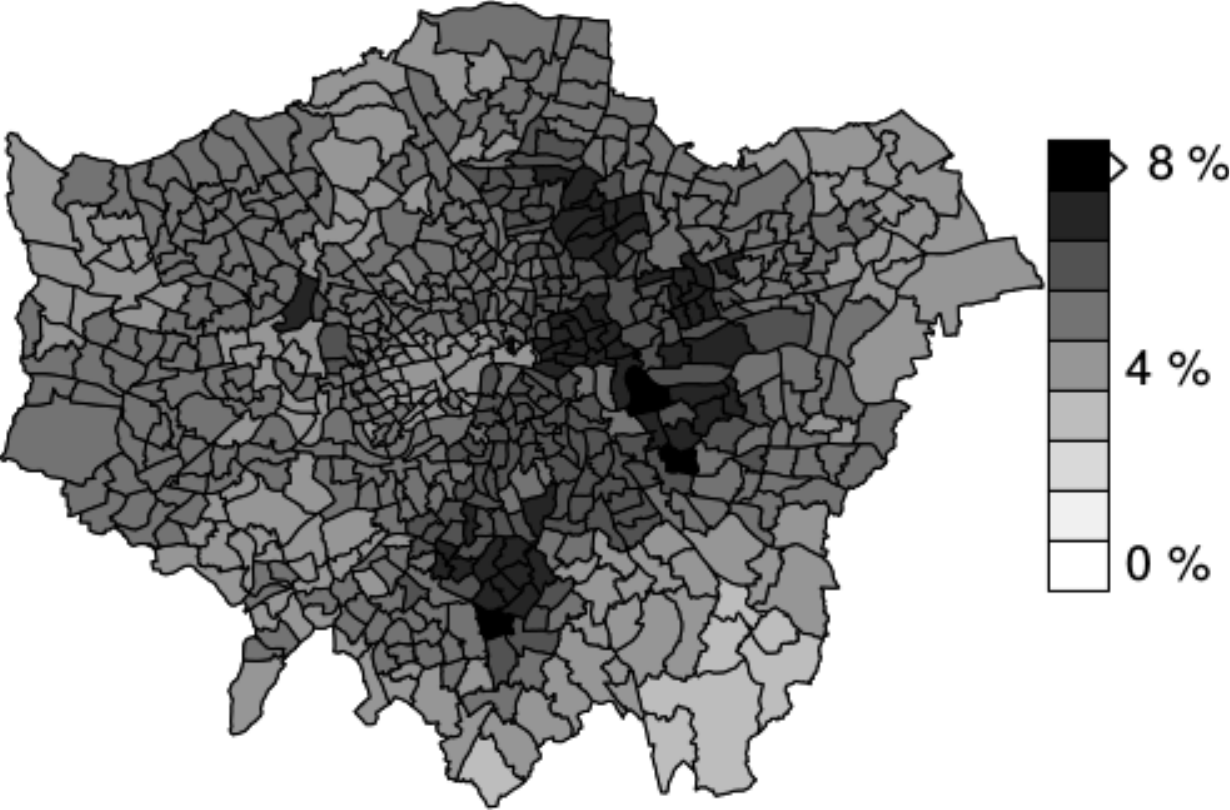} \\
(a) & (b)\\
\includegraphics[width=0.3\textwidth]{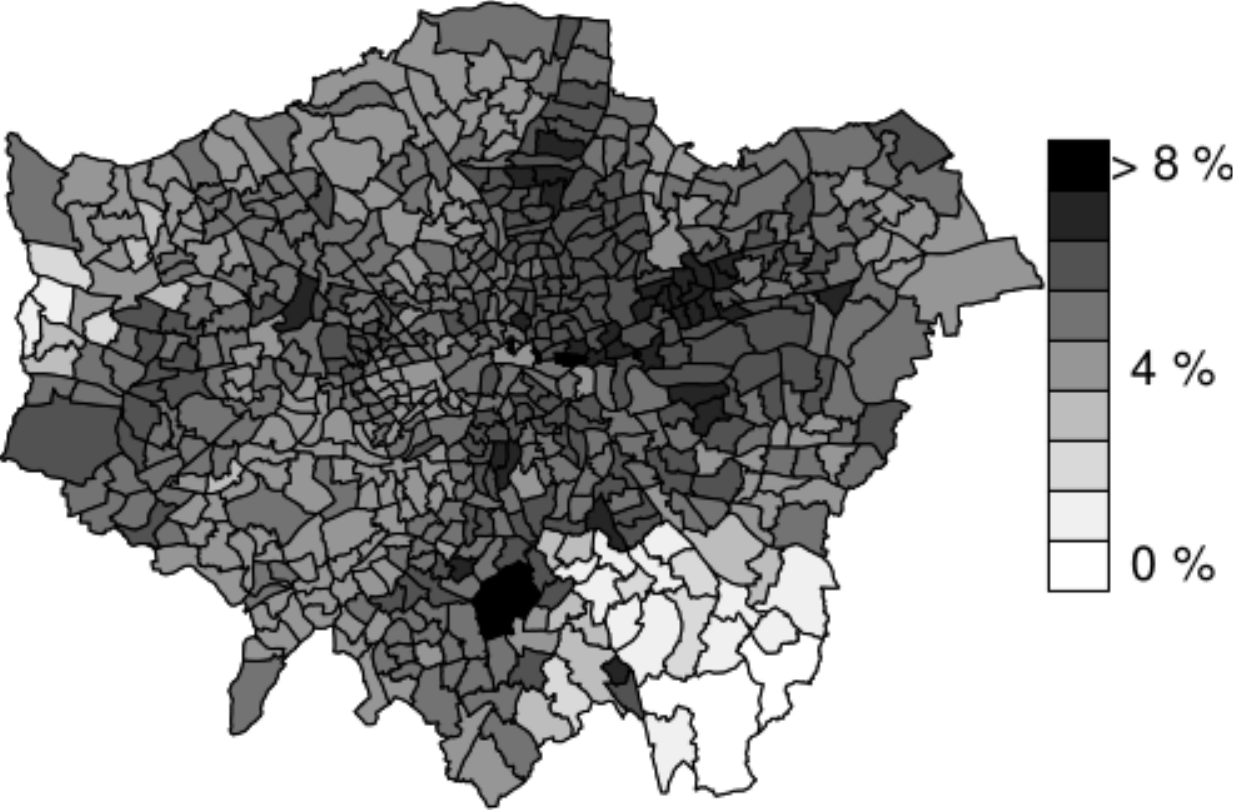} &
\includegraphics[width=0.3\textwidth]{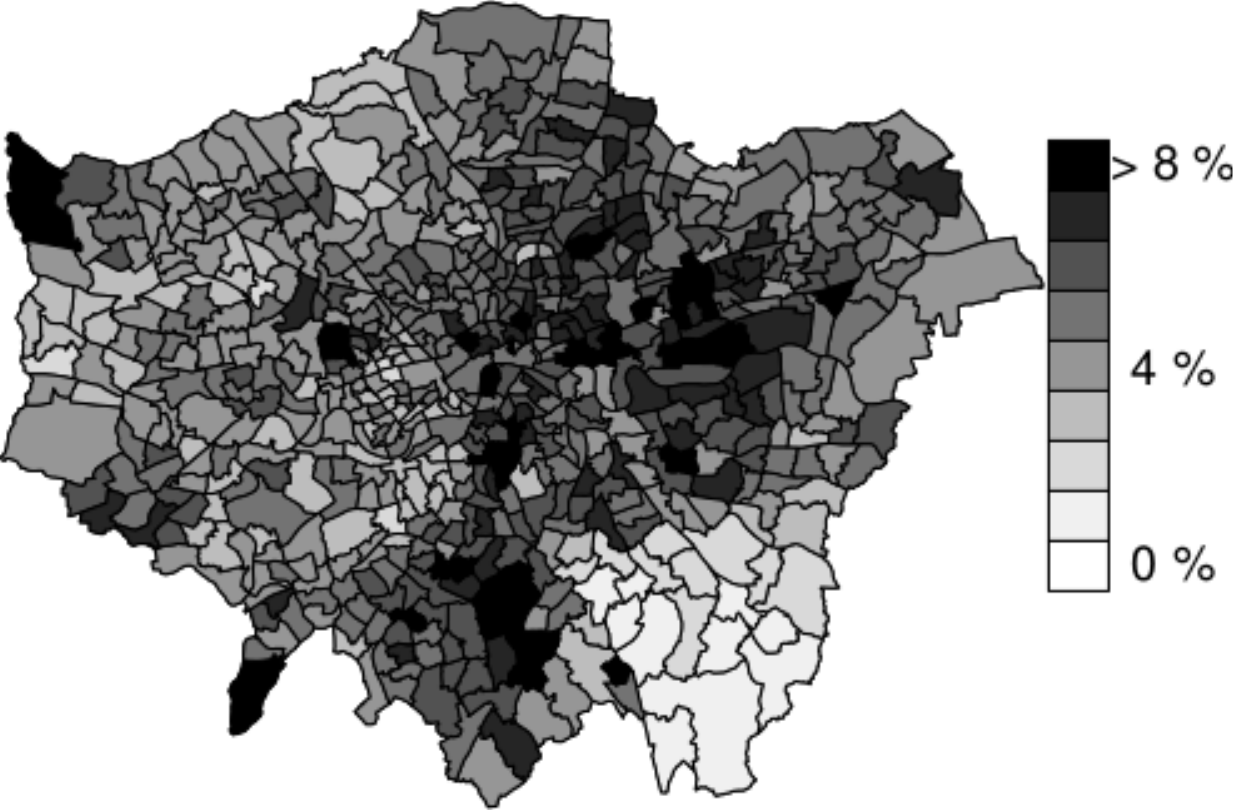}\\
(c) & (d)\\
\end{tabular}
\end{center}
\caption{Probabilities of preterm birth: (a) observed, and smoothed utilizing (b) the proposed model, (c) the model of
\citet{FG2002}, and (d) an MCAR model.}\label{app.ptb}
\end{figure}

\begin{figure}
\begin{center}
\begin{tabular}{cc}
\includegraphics[width=0.3\textwidth]{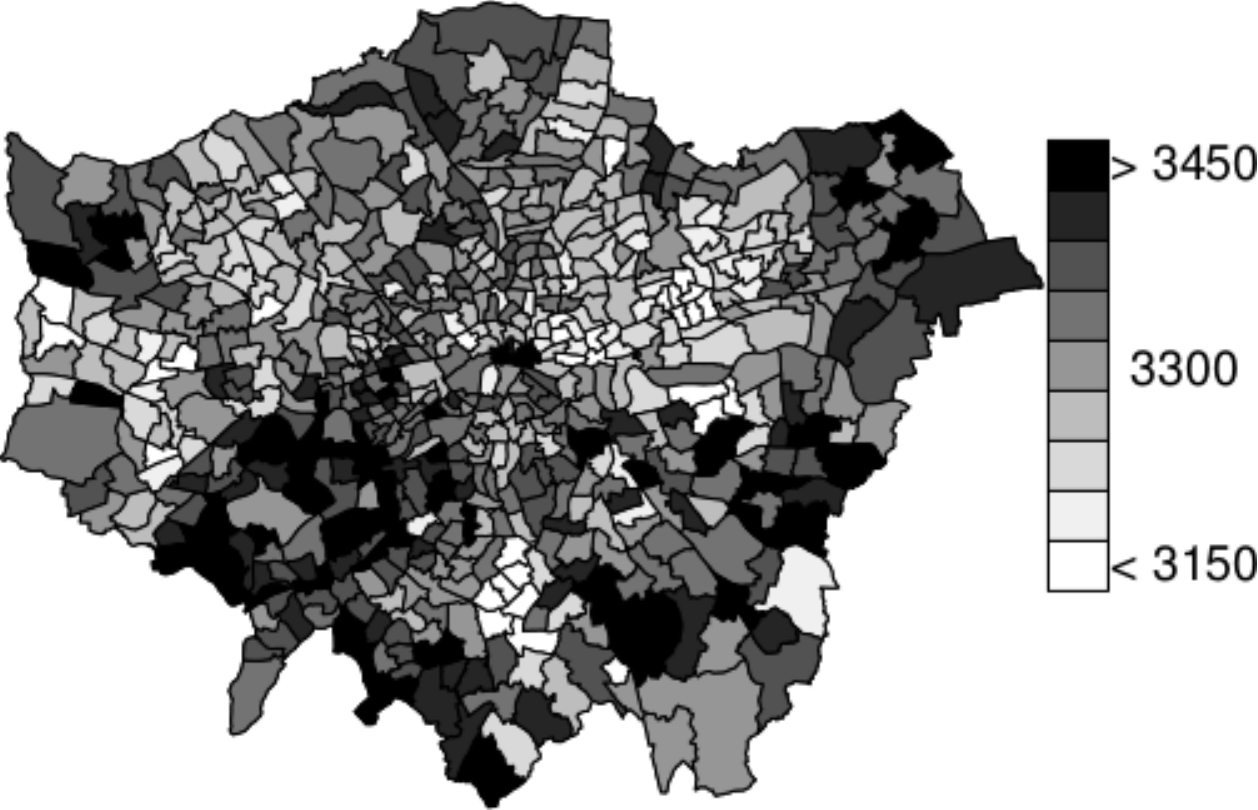} &
\includegraphics[width=0.3\textwidth]{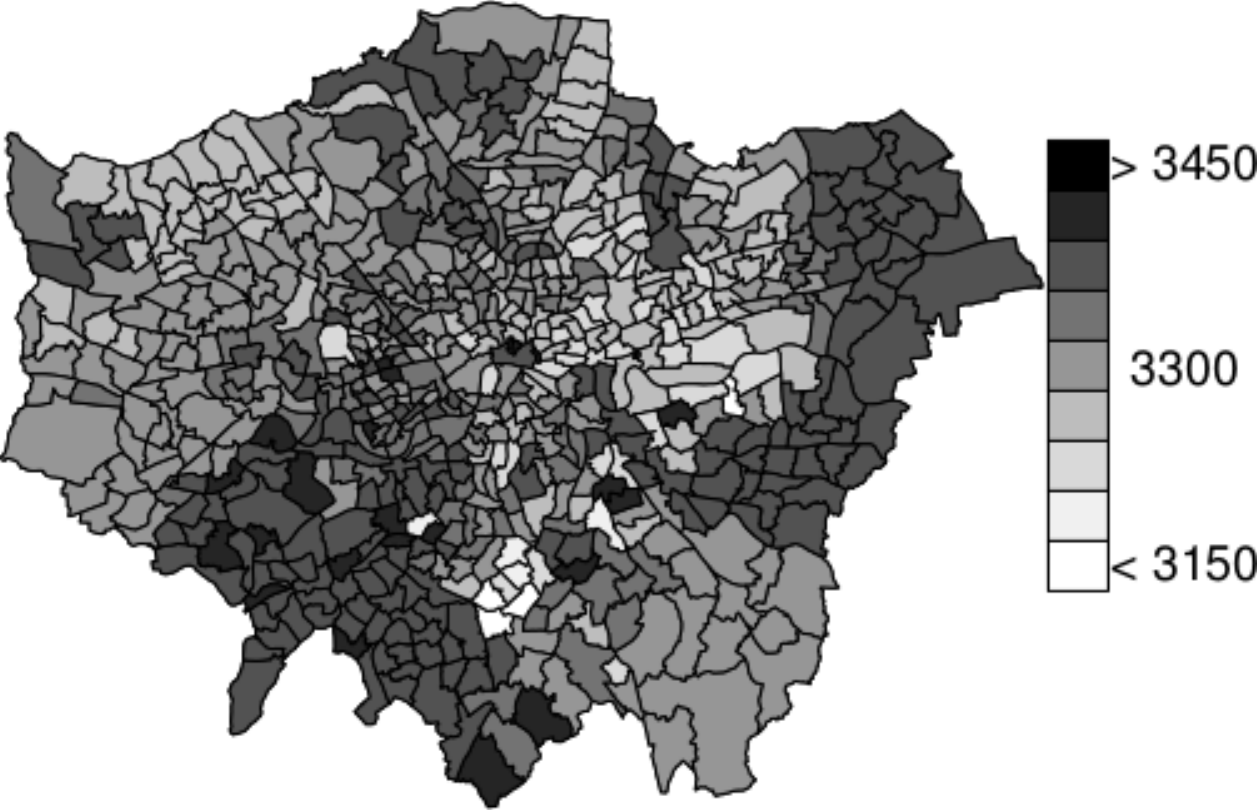} \\
(a) & (b)\\ 
\includegraphics[width=0.3\textwidth]{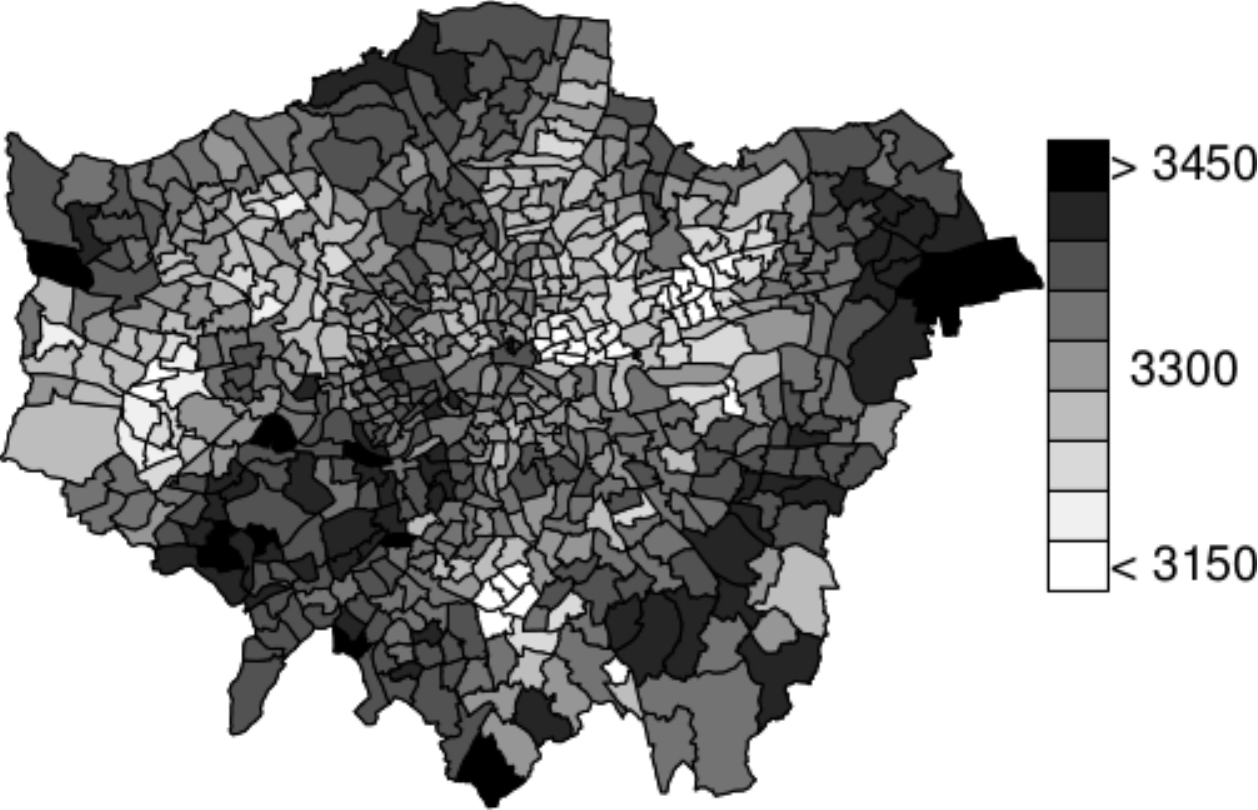} &
\includegraphics[width=0.3\textwidth]{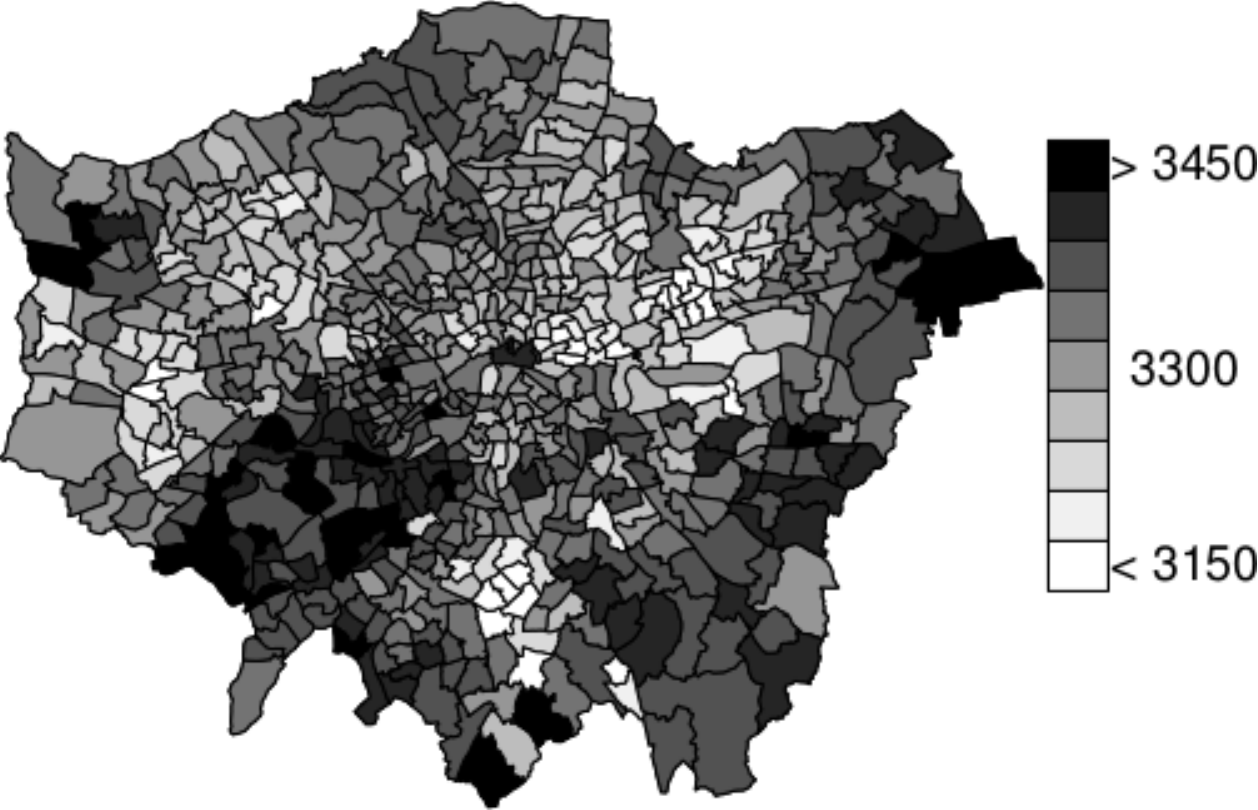} \\
(c) & (d)\\
\end{tabular}
\end{center}
\caption{Birth weight: (a) observed, and smoothed utilizing (b) the proposed model, (c) the model of
\citet{FG2002}, and (d) an MCAR model.}\label{app.wgh}
\end{figure}

\begin{figure}
\begin{center}
\begin{tabular}{cc}
\includegraphics[width=0.3\textwidth]{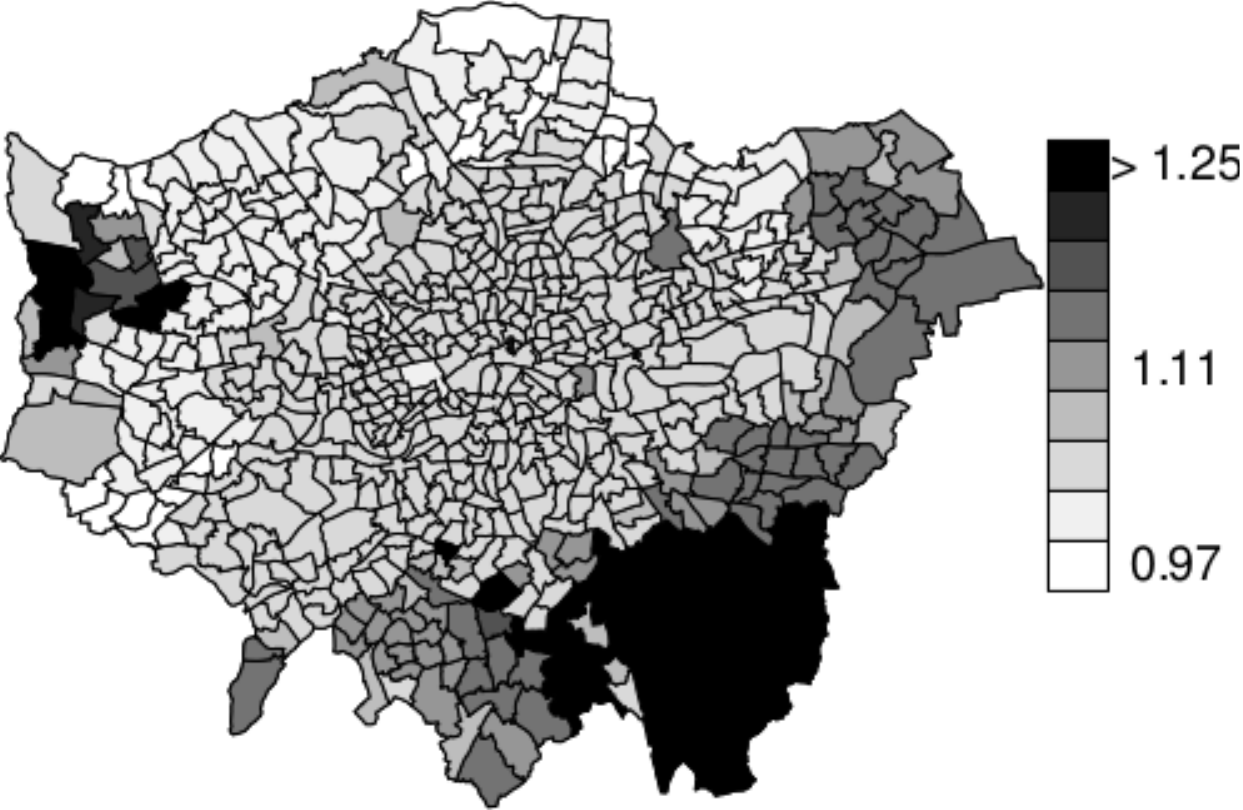} &
\includegraphics[width=0.3\textwidth]{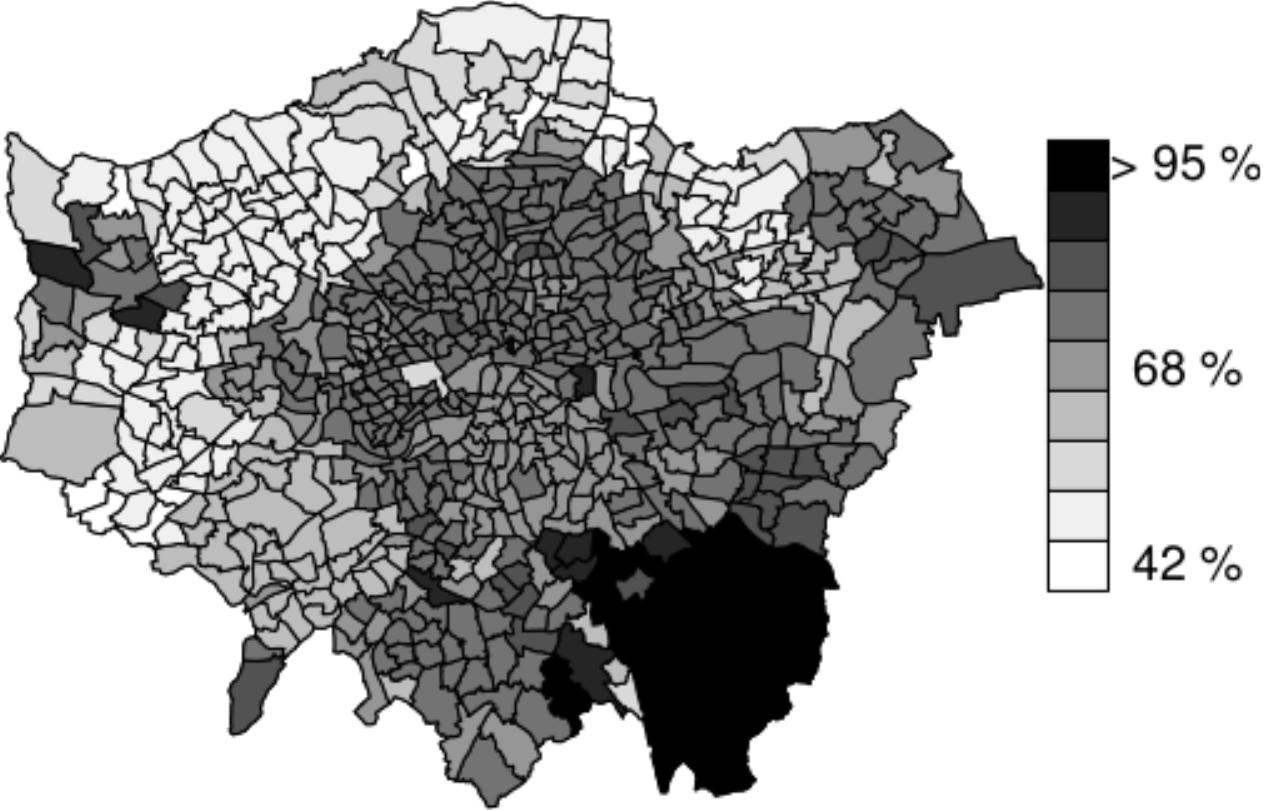}\\
(a) & (b)\\
\includegraphics[width=0.3\textwidth]{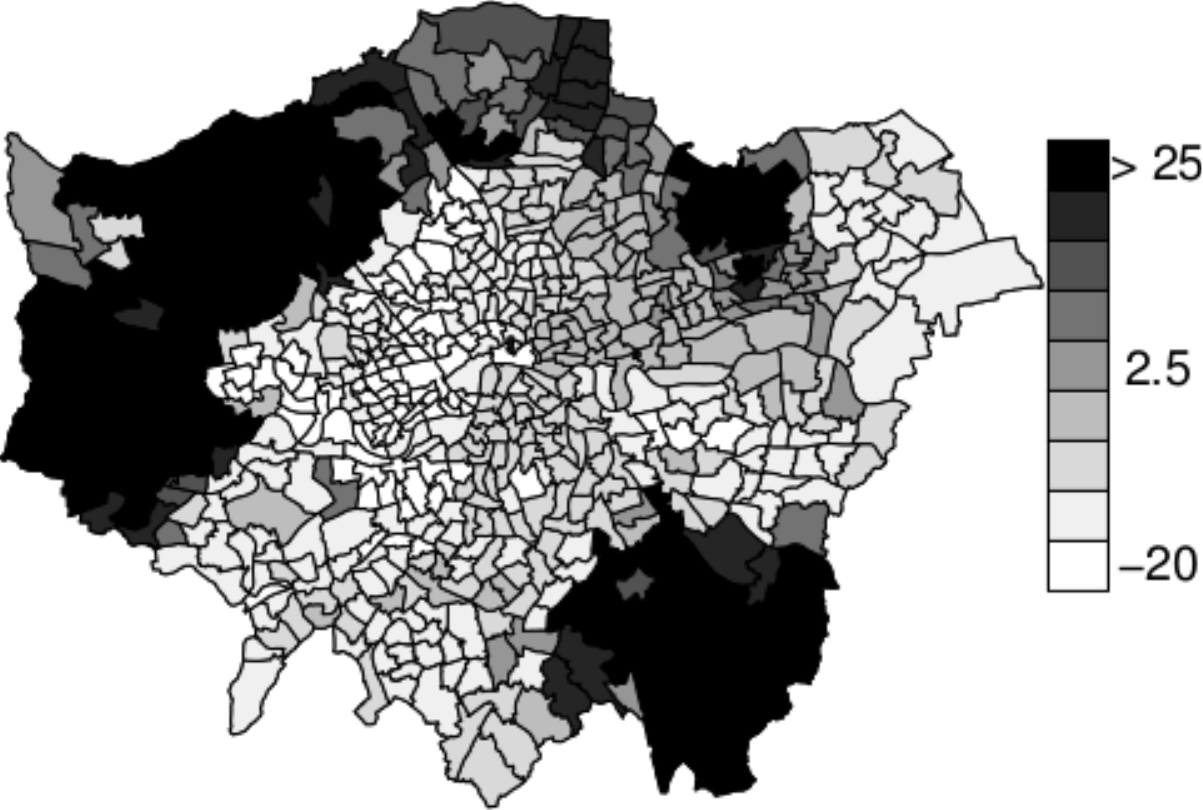} &
\includegraphics[width=0.3\textwidth]{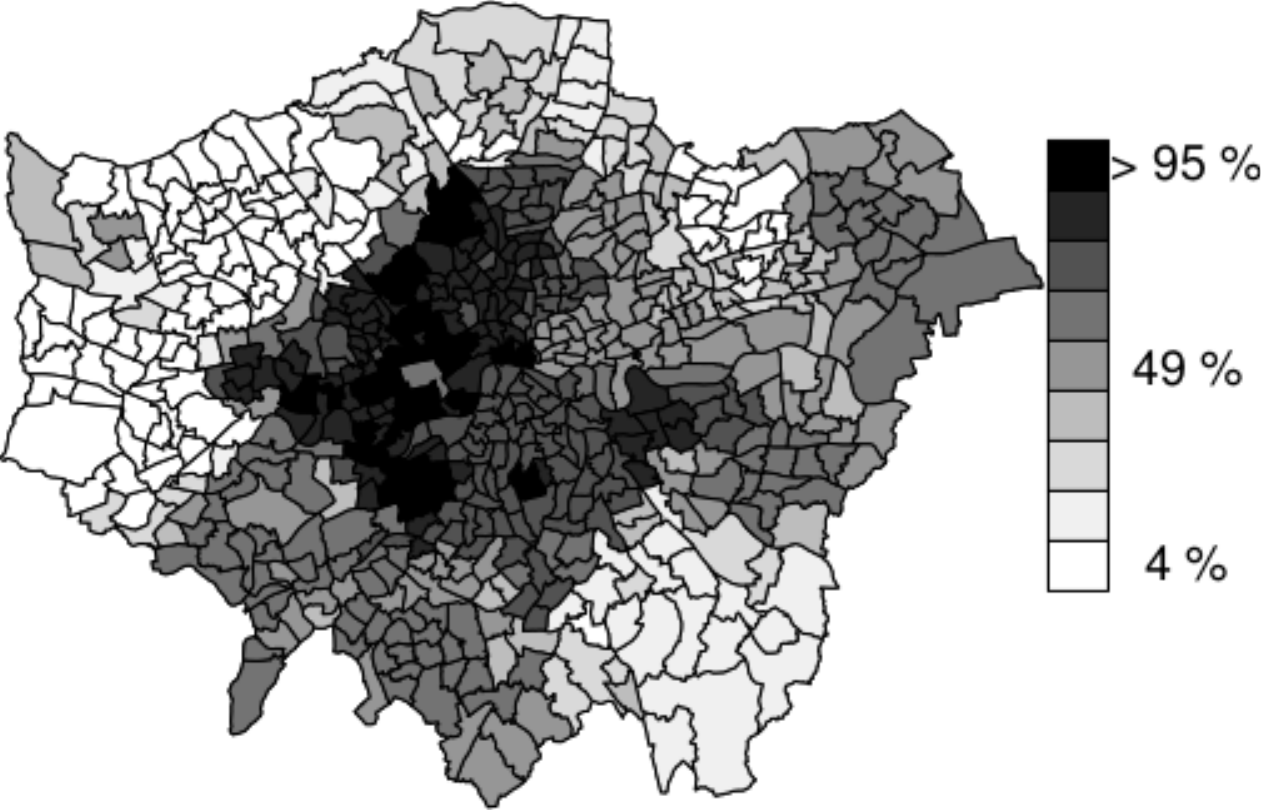}\\
(c) & (d)\\
\end{tabular}
\end{center}
\caption{
(a) Posterior medians of $\exp(\beta_{i,11})$ which represent odds ratio of preterm birth when increasing exposure to PM$_{10}$ by one
- the interquartile range of PM$_{10}$. 
(b) Posterior probabilities that $\beta_{i,11}$ are bigger that zero, $P(\beta_{i,11} > 0 |\text{data})$.    
(c) Posterior medians of $\beta_{i,12}$, the coefficients of exposure to PM$_{10}$ in the model for birth weight.
(d) Posterior probabilities that $\beta_{i,12}$ are less that zero, $P(\beta_{i,12} < 0 |\text{data})$.
}\label{app.res2}
\end{figure}

The model we fit takes the form
$f_{i}(\uy_{i},\uw_{i}|\ux_{i}) = \sum_{h=1}^{\infty} \pi_{hi} f(\uy_{i},\uw_{i}|\ux_{i};\utheta_h)$.
It is a joint model of two discrete variables, the Binomial $Y_{i1}$ and count $W_{i1}$, and eight continuous ones,
$Y_{i2}$ and $W_{i2}, \dots, W_{i8}$. Only the means of the response variables, $Y_{i1}, Y_{i2},$ are modeled in terms of
explanatory variables, $\ux_{i1}=(1,\text{PM}_{10,i})^T$ and $\ux_{i2}=(1,\text{PM}_{10,i},\text{O}_{i})^T$ respectively. Further, variables $W_{ij}, j=1,\dots,8,$ 
are jointly modeled with the responses in order to adjust for their effects. Data and results are displayed in Figures \ref{pm10} - \ref{app.res2}.

First, Figure \ref{pm10} displays the estimated average annual exposures to PM$_{10}$. 
These range from 17.1 to 22 $\mu g /m^3$, with average exposure equal to 18.6 $\mu g /m^3$,
and interquartile range of 1 $\mu g /m^3$.

Figure \ref{app.ptb} (a) displays the observed area-wise probabilities of preterm birth.  
Smooth estimates recovered from the proposed model are displayed in Figure \ref{app.ptb} (b). 
They are obtained as follows: for each area $i, i=1\dots,n,$ and for each iteration of the sampler $t, t=1,\dots,T$, we   
observe the cluster assignment and the regression coefficients associated with this cluster. Denote these by 
$z_i^{(t)}$ and $\ubeta_{i1}^{(t)}$ respectively, where $\ubeta_{i1}^{(t)}$ depends on $i$ through $z_i^{(t)}$. 
The model based estimate of the probability of preterm birth in area $i$ at iteration $t$ is obtained as 
$\pi_i^{(t)} = \text{logit}^{-1}(\ux_{i1}^T \ubeta_{i1}^{(t)})$, while the smooth model based estimate 
of the same probability is obtained as median$(\pi_{i}^{(1)},\dots,\pi_{i}^{(T)})$. 

For a comparison, we also obtained smooth model based estimates of the probabilities of preterm birth by generalizing the
model proposed by \citet{FG2002} to handle mixed type outcomes. These are shown in Figure \ref{app.ptb} (c).
We have also fitted a multivariate generalized linear mixed model with random effects that have multivariate 
conditionally autoregressive (MCAR) distributions (see e.g. \citet{Mardia}, \citet{Gelfand2003}, \citet{Jin2005}). 
These are shown in Figure \ref{app.ptb} (d).
Briefly, the model of \citet{FG2002} here is expressed as
$f_{i}(\uy_{i}|\ux_{i},\uw_{i}) = \sum_{h=1}^{\infty} \pi_{hi} f(\uy_{i}|\ux_{i},\uw_{i};\utheta_h)$.
It is a joint model of two response variables, the binomial $Y_{i1}$ and the continuous $Y_{i2}$,
where the corresponding latent and observed continuous variables, $y^*_{i1}$ and $y_{i2}$, are modeled as  
\begin{eqnarray}
&&y^*_{i1} = \ux^T_{i1} \ubeta_{i,11} + \uw^T_{i} \ubeta_{i,21} + \epsilon_{i1} \nonumber\\ 
&&y_{i2} = \ux^T_{i2} \ubeta_{i,12} + \uw^T_{i} \ubeta_{i,22} + \epsilon_{i2}, \nonumber 
\end{eqnarray}
and the bivariate error term is assumed to be distributed as
\begin{eqnarray}
\left( 
\begin{array}{c}
\epsilon_{1i} \\ 
\epsilon_{2i}
\end{array}
\right) \stackrel{iid}{\sim} N_2 \left( \left[ 
\begin{array}{c}
0 \\ 
0
\end{array}
\right] ,\left[ 
\begin{array}{cc}
1 & \sigma_{12} \\ 
\sigma_{21} & \sigma_{22}
\end{array}
\right] \right). \nonumber
\end{eqnarray}

Continuing now with the MCAR model, it is expressed as
\begin{eqnarray}
Y_{i1} &\sim& \text{Binomial}(N_i,\pi_i) \nonumber\\ 
\text{logit}(\pi_i) &=& \ux^T_{i1} \ubeta_{11} + \uw^T_{i} \ubeta_{21} + \ux^T_{i1} \ubeta_{i,11} + \uw^T_{i} \ubeta_{i,21} \nonumber\\
Y_{i2}              &=& \ux^T_{i2} \ubeta_{12} + \uw^T_{i} \ubeta_{22} + \ux^T_{i2} \ubeta_{i,12} + \uw^T_{i} \ubeta_{i,22} + \epsilon_{i}, \nonumber 
\end{eqnarray}
where $\{\ubeta_{i,11}, \ubeta_{i,21}, \ubeta_{i,12}, \ubeta_{i,22}: i=1,\dots,n\}$ denote area-specific random effects.
For random effects that appear in only one of the response models, a univariate CAR model 
is assumed, while for effects that appear in both, a bivariate CAR model is specified.
For instance, the observed proportion of preterm births $O_i$ is included only in the model of birth weight $Y_{i2}$,
and the corresponding random effects $\ubeta^{*} = (\beta^*_1,\dots,\beta^*_n)$ are modeled as
\begin{eqnarray}
 \beta_{j}^*| \beta_{-j}^*, \tau^2 \sim N(\sum_{i \sim j} \beta^*_i/n_j, \tau^2/n_j),\nonumber
\end{eqnarray}
where $n_j$ denotes the number of neighbors of area $j$, and $\tau^2$ is a variance parameter. 

All other random effects are independently modeled using bivariate CAR distributions. For instance, random effects  
corresponding to PM$_{10}$, $\ubeta_{i,11}=(\beta_{i,111}, \beta_{i,121})^T, i=1,\dots,n,$ 
\begin{eqnarray}
 \ubeta_{j,11}| \ubeta_{-j,11}, \uV \sim N_2(\sum_{i \sim j} \ubeta_{i,11}/n_j, \uV/n_j),\nonumber
\end{eqnarray}
where $\uV$ is a $2 \times 2$ positive definite matrix.

By comparing Figure \ref{app.ptb} (b) to (c) and (d), it appears that the proposed model
results in smoother estimates of the preterm birth probabilities. This can be attributed to the fact that the proposed model 
fits a simple response model within each component. This results in partitions with higher numbers of active, i.e nonempty, components,
and such partitions are characterized by higher levels of uncertainty. A similar observation about the degree of smoothness can be made 
in Figure \ref{app.wgh} which displays the data and model based estimates of birth weight. 

Lastly, Figure \ref{app.res2} examines the effects of PM$_{10}$ on the probability of preterm birth and 
birth weight. Figure \ref{app.res2} (a) displays the posterior medians of the area-wise log odds ratios of preterm birth 
when increasing the exposure to PM$_{10}$ by one - the interquartile range of PM$_{10}$. 
These are obtained by averaging over all 
iterations the area specific odds ratios, $\exp(\beta_{i,11})$.
We see that the model identifies a cluster in the SE and a smaller one in the NW with higher 
odds of preterm birth. Estimates based on the model of 
\citet{FG2002} and the MCAR model exhibit similar behaviors, and hence corresponding results are not displayed. 
Figure \ref{app.res2} (b) displays the posterior probabilities that $\beta_{i,11}$ are larger than zero:
P$(\beta_{i,11}>0|\text{data})$. These are higher than $95\%$ over the two aforementioned clusters of areas. 
Further, Figure \ref{app.res2} (c) displays the posterior means of $\beta_{i,12}$. These describe 
the estimated effect of increasing exposure to PM$_{10}$ by 1 $\mu g /m^3$ on birth weight. Figure \ref{app.res2} (d) displays the posterior probabilities that $\beta_{i,12}$ are less than zero:
P$(\beta_{i,12}<0|\text{data})$. We see that the model identifies a group of areas in the central part of London
for which P$(\beta_{i,12}<0|\text{data})>0.95$. The corresponding estimated effects have posterior means not less than $-20$g. 

\section{Discussion}\label{discussion}

We have developed Bayesian nonparametric models for spatially distributed data of mixed type that 
aim at providing a flexible way of adjusting for the effects of confounding variables and hence allowing
for efficient estimation of the regression coefficients of interest. We have compared the proposed
model, $f_{i}(\uy_{i},\uw_{i}|\ux_{i}) = \sum_{h=1}^{\infty} \pi_{hi} f(\uy_{i},\uw_{i}|\ux_{i};\utheta_h)$, 
to models of the form $f_{i}(\uy_{i}|\uw_{i},\ux_{i}) = \sum_{h=1}^{\infty} \pi_{hi} f(\uy_{i}|\uw_{i},\ux_{i};\utheta_h^*)$ 
\citep{FG2002, GR2002},
the more recent ones that take the form $f_{i}(\uy_{i},\uw_{i},\ux_{i}) = \sum_{h=1}^{\infty} \pi_{hi} 
g(\uy_i|\ux_i,\uw_i;\utheta_h^{'})h(\ux_i,\uw_i;\utheta_h^{'})$ 
\citep{Shahbaba, Hannah}, some special cases of the those and the more classical (M)CAR \citep{BYM,Mardia} models. 
Our simulation studies
have shown situations in which the proposed model can do well in terms of estimating the underlying
regression coefficients. 

Computationally, the model we have proposed can be quite demanding, depending of course 
on the dimension and type of the variables included. 
There are two main steps in the MCMC algorithm, other than the updating of the GMRFs that is common to
all models, that can be computationally intensive. Firstly, the numerical integration
over the unobserved latent variables is numerically intensive and it can create numerical problems when 
integrating over latent variable distributions that correspond to empty clusters, as these 
sometimes can have covariance matrices that are close to being singular. However, we have chosen to perform 
the integration, instead of imputing the latent variables, as this greatly improves the mixing
of the algorithm. Secondly, joint modeling 
of multivariate responses and confounders creates the need of handling possibly high dimensional 
covariance and precision matrices, which is also computationally demanding.
Alternatively, one could impose diagonal covariance matrices, as in model M$_3$ that we examined in the simulation
study, but this option, in the scenario we examined, was not the best one in terms of RAMSE.  

The possible high dimensionality of the vector of responses and confounders and the computational problems it
creates can potentially be alleviated by developing a variable selection algorithm that excludes from the model 
confounding variables that create spurious clusters. In addition, a variable selection algorithm that excludes from the
within cluster regression model risk factors that do not have an effect on the risk of the cluster 
is also of interest. Consider for instance a case similar to the one presented in the simulation studies, 
that is, a case where there is one count response variable, one risk factor, and one confounding variable.  
Further suppose that for most of the iterations of the sampler, two cluster are identified,
for which linear predictors of the form $\eta_h = \beta_{0h} + \beta_{1h}x, h=1, 2,$ adequately
describe the within cluster risk-risk factor relationship. Introduction now of a 
second confounding variable that contains no relevant information can potentially split the clusters into 
smaller ones. Of course, this will have a negative effect on the estimation of the within cluster
regression coefficients. For instance, the second confounding variable could be a `coin flip', 
meaning a binary variable that carries no information, that will split each of the two legitimate clusters into 
two smaller ones. Denote the new linear predictors as 
$\eta_{hk} = \beta_{0hk} + \beta_{1hk} x, h=1, 2, k=1, 2$. Under this scenario, hypothesis tests of the form 
H$_0$: $\beta_{0h1} = \beta_{0h2}$, and H$_0$: $\beta_{1h1} = \beta_{1h2}$, $h=1, 2$, 
will not be rejected with high probability. This can be the basis of a variable selection
algorithm suitable for the proposed model. Furthermore, continuing on the same example, exclusion of the 
within cluster risk factor can be performed in a straight forward way, based  on hypothesis tests of the form 
H$_0$: $\beta_{1h} = 0$. 

\section{Acknowledgements}
\vspace{-0.1cm}
The authors thank the Medical Research Council (MRC) (grant number G09018401) for partially funding this research project,
the Small Area Health Statistics Unit (SAHSU) and Anna Hansell of SAHSU, Imperial College London, for providing health, population,
and birth data from Hospital Episode Statistics (HES) of the  Health and Social Care Information Centre (HSCIC), 
and the Environmental Research Group, King's College London, for providing the annual average exposure estimates of PM$_{10}$. 
HES data are copyright \textcopyright 2013, re-used with the permission of HSCIC. All rights reserved. 
The population and cancer data were supplied to SAHSU by the Office for National Statistics, derived from national cancer 
registrations and the Census. Data providing organizations did not participate in analysis or writing of this manuscript.
Special thanks are due to Alex Beskos of University College London for his insightful discussion on the development 
of the MCMC sampler, and two anonymous referees for their insightful comments that have substantially improved this paper. 

\section{Appendix: MCMC algorithm}\label{appendix}

Our sampler utilizes the following steps:
\begin{enumerate}

\item Update $\uxi_{h}, h \geq 1,$ from 
\begin{eqnarray}\nonumber
\uxi_{h} | \dots \sim N_{r_3+q}\left(
\uB \left\{\sum_{i:\delta_i=h} \uX_i^{*^T}\uSigma_h^{*^{-1}} \uv_i+\uD_{\xi}^{-1}\umu_{\xi}\right\},
\uB \equiv \left\{ \sum_{i:\delta_i=h} \uX_i^{*^T} \uSigma_h^{*^{-1}}\uX_i^{*}+\uD_{\xi}^{-1} \right\}^{-1}
\right).
\end{eqnarray}

\item To sample from the posterior of the restricted covariance matrix $\uSigma_h^{*}, h \geq 1,$ we use 
the parameter-extended algorithm of \citet{xiao} that requires the joint posterior of $(\uD_h, \uSigma_{h}^{*})$.
This, apart from a normalizing constant, is given by
\begin{eqnarray}\nonumber
 p(\uD_h, \uSigma_{h}^{*}|\dots) \propto |\uD_h|^{\eta/2-1} |\uSigma_{h}^*|^{(\eta-s-1-n_h)/2} 
\text{etr}\{-(\uEta^{-1} \uE_{h}+\uSigma_{h}^{*^{-1}} \uS_{h})/2\},
\end{eqnarray}  
where $\uS_h = \sum_{i:\delta_i=h}(\uv_i-\umu_i^*)(\uv_i-\umu_i^*)^T$.

Sampling at iteration $t+1$ proceeds as follows: given realizations from iteration $t$, 
$\uD_h^{(t)}$, $\uSigma_{h}^{*^{(t)}}$, we propose new values by generating 
$\uE_h^{(p)}\sim$  
Wishart$_s(\uE_h^{(p)};\psi,\uE_h^{(t)}/\psi)$.
Here, $\uE_h^{(t)}=\uD_h^{(t)^{1/2}} \uSigma_h^{*^{(t)}} \uD_h^{(t)^{1/2}}$, and proposed values are obtained by decomposing
$\uE_h^{(p)}=\uD_h^{(p)^{1/2}} \uSigma_h^{*^{(p)}} \uD_h^{(p)^{1/2}}$.
Proposed values are accepted with probability 
$$\alpha = \min\left\{\frac{p(\uD_h^{(p)}, \uSigma_{h}^{*^{(p)}}|\dots)}{p(\uD_h^{(t)}, \uSigma_{h}^{*^{(t)}}|\dots)}
\frac{t(\uD_h^{(t)}, \uSigma_{h}^{*^{(t)}}|\uD_h^{(p)}, \uSigma_{h}^{*^{(p)}})}
{t(\uD_h^{(p)},\uSigma_{h}^{*^{(p)}}|\uD_h^{(t)}, \uSigma_{h}^{*^{(t)}})},1\right\},$$
where, the proposal density is given by
$t(\uD_h^{(p)},\uSigma_{h}^{*^{(p)}}|\uD_h^{(t)}, \uSigma_{h}^{*^{(t)}})
= \text{Wishart}_s(\uE_h^{(p)};\psi,\uE_h^{(t)}/\psi)J(\uE_{h}^{(p)} \rightarrow \uD_h^{(p)}, \uSigma_{h}^{*^{(p)}})$.
We choose the degrees of freedom $\psi$ so as to achieve an acceptance ratio of about $20 - 25\%$ \citep{Roberts2001c}.

\item Vectors of regression coefficients $\ubeta_{h,1:2} = (\ubeta_{h1}^T, \ubeta_{h2}^T)^T$, $h \geq 1$, 
are updated from the marginal posterior, having integrated out $\uy_{i,1:2}^*$.
We first partition $\uv_i = ((\uy^{*}_{i})^T, \uw_{i}^T)^T$ into unobserved and observed variables 
$\uv_i = (\uy_{i,1:2}^{*^T},\us_i^T)$. The distribution of $\uv_i$, given in (\ref{wis}), is now re-written as 
\begin{eqnarray}\nonumber
\uv_{i} |(\umu_i^*,\uSigma^*_i)  \sim N_s\left(
\umu_i^*=
\begin{array}{cc}
\left( 
\begin{array}{l}
0\\
0\\
\umu_{i,s} \\
\end{array}
\right),
 &
\uSigma^*_i=\left[ 
\begin{array}{ll}
\uR_i &  \uF_{i} \\
\uF_{i}^T & \uG_i\\
\end{array}
\right]
\end{array}\right).
\end{eqnarray}
The regression coefficients are updated from $p(\ubeta_{h,1:2}|\dots) \propto$
\begin{eqnarray}
\prod_{\{i:\delta_i=h\}}\Big[\int_{\Omega_{i2}} \int_{\Omega_{i1}} N_2\{\uy_{i,1:2}^* | 
\uF_h \uG_h^{-1}(\us_i-\umu_{i,s}),\uR_h-\uF_h\uG_h^{-1}\uF_h^T\}d\uy_{i,1:2}^*
\Big] N(\ubeta_{h1},\ubeta_{h2};\uzero,\tau^2\uI),\nonumber
\end{eqnarray}
where $\Omega_{ik} = (c_{i,k,y_{i1}-1}, c_{i,k,y_{i1}}), k=1,2$.

At iteration $t+1$, utilizing the realization from the previous iteration 
$\ubeta_{h,1:2}^{(t)}$, we propose a new value: $\ubeta_{h,1:2}^{(p)} = \ubeta_{h,1:2}^{(t)} + \uepsilon$, where 
$\uepsilon \sim N_{2r}(\uzero,\tau^2_{\epsilon} \uI)$. We choose $\tau^2_{\epsilon}$ in order to achieve acceptance rate 
about $20 - 25\%$ \citep{Roberts2001c}. 
The proposed value is accepted with probability 
$\alpha = \min\{p(\ubeta_{h,1:2}^{(p)}|\dots)/p(\ubeta_{h,1:2}^{(t)}|\dots),1\}$.

\item Impute latent vectors $\uy_{i,1:2}^*, i=1,\dots,n,$ from 
$$\uy_{i,1:2}^{*} \sim N_2(\uF_h \uG_h^{-1}(\us_i-\umu_{i,s}),\uR_h-\uF_h\uG_h^{-1}\uF_h^T)
I[y_{i1}^{*} \in \Omega_{i1}] I[y_{i2}^{*} \in \Omega_{i2}]$$
using the algorithm described by \citet{Robert2009}, that is, by imputing one element 
of $\uy_{i,1:2}^{*}$ at a time given the other one.

\item Update the allocation variables $\delta_{i},i=1,2,\dots,n,$ according to allocation probabilities
obtained from the marginalized posterior
\begin{eqnarray}\nonumber
P(\delta_{i}=h) \propto
\Big[\int_{\Omega_{i2}} \int_{\Omega_{i1}} N_2\{\uy_{i,1:2}^* | 
\uF_h \uG_h^{-1}(\us_i-\umu_{i,s}),\uR_h-\uF_h\uG_h^{-1}\uF_h^T\}d\uy_{i,1:2}^*\Big]
N_{1+q}(\us_i|\umu_{i,s},\uG_{h})\pi_{hi}.
\end{eqnarray}

\item Label switching moves (a generalization from \citet{PR08}): 
\begin{enumerate}

\item[(a)] Propose to change the labels $a$ and $b$ of two randomly chosen nonempty components.
The proposed change is accepted with probability 
$\min \left(1,\prod_{i:\delta_{i}=a}\frac{\pi_{bi}}{\pi_{ai}} \prod_{i:\delta_{i}=b}\frac{\pi_{ai}}{\pi_{bi}}\right)$.
If the proposed swap is accepted, change allocation variables and cluster specific parameters.

\item[(b)] Propose to change the labels $a$ and $a+1$ of two components, but at the same time propose to 
exchange $\ueta_a$ with $\ueta_{a+1}$, where $\ueta_h = (\eta_{h1},\dots,\eta_{hn})^T$. 
Cluster with label $a$ is chosen uniformly among all components labeled
$1,\dots,n^*-1$, where $n^*$ is the nonempty component with the largest label.
The proposed move is accepted with probability \\
$\min \left(1,\prod_{i:\delta_{i}=a} \{1-\Phi(\eta_{a+1,i})\} \prod_{i:\delta_{i}=a+1} 
\{1-\Phi(\eta_{ai})\}^{-1} \right)$. If the proposed swap is accepted, change allocation variables
and cluster specific parameters. 

\end{enumerate}
\end{enumerate}

To update the Gaussian Markov random fields and subsequently parameters $(\alpha,\phi,\lambda)$,
we first note that $\ueta_h \stackrel{\text{iid}}{\sim} N_n(\alpha \mathbf1_n, \phi^{-2} \uQ_{\lambda}^{-1})$. 
Further, we introduce independent latent 
variables $z_{hi} \sim N(\eta_{hi},1)$ and 
define $\delta_{i}=k_{i}$ if and only if $z_{li} > 0$ for $l=k_{i}$ and 
$z_{li} < 0$ for $l<k_{i}$. A very similar augmentation scheme was proposed by 
\citet{RD11}. Our approach differs from that of \citet{RD11} in that we augment with
$\{z_{li}\}_{l=1}^{k_{i}}$ whereas \citet{RD11} augment with $\{z_{li}\}_{l=1}^{T}$.
In the \citet{RD11} approach variables $\{z_{li}\}_{l=k_{i}+1}^T$ are imputed from the 
prior as there is no information in the data about these, resulting in samples from the 
posteriors of $(\alpha,\phi,\lambda)$ in which the prior receives excess weight.  
A drawback of our approach, however, as becomes clear in the following updating steps,  
is that it is more involved and computationally demanding.  

The corresponding complete data likelihood is
\begin{eqnarray}
&&\ell\left(\left\{\uy_{i},\uw_{i},\delta_{i}=k_{i}, \{z_{li}\}_{l=1}^{k_{i}}: i=1,\dots,n\right\}\right)=\nonumber\\
&& \prod_{i} \left\{ f(\uy_{i},\uw_{i}|\ux_{i};\utheta_{k_{i}}) P(\delta_{i}=k_{i}|z_{li}: l \leq k_{i})
d(z_{li}: l \leq k_{i})\right\}=\nonumber\\
&& \prod_{i} \Big\{ f(\uy_{i},\uw_{i}|\ux_{i};\utheta_{k_{i}})
I[z_{k_{i}i} > 0 \text{\;and\;} z_{li}<0 \text{\;for\;} l < k_{i}]
\prod_{l=1}^{k_{i}} N(z_{li};\eta_{li},1)\Big\}.\nonumber
\end{eqnarray}
The sampler updates from  
$\pi(\utheta,\udelta,\uz,\ueta,\alpha,\phi,\lambda,\uy^*|\uy,\uw) \propto
h_1(\udelta|\uz) h_2(\uz|\ueta) h_3(\ueta,\alpha,\phi,\lambda)$ 
as follows 

\begin{enumerate}

\item[7.] For $h < k_{i}$ we update 
$z_{hi} \sim N(\eta_{hi},1) I[z_{hi} < 0]$, and for 
$h = k_{i}$ we update     
$z_{hi} \sim N(\eta_{hi},1) I[z_{hi} > 0]$.

\end{enumerate}

We now obtain a sample from $\pi(\ueta,\alpha,\phi,\lambda|\dots) = 
\pi(\ueta^{(A)},\ueta^{(D)},\alpha,\phi,\lambda|\uz)$. 
Here $\ueta=\{\ueta_h: h=1,2,\dots\}$ and $\ueta^{(A)}=\{\ueta_h^{(A)}: h=1,2,\dots\},$ where 
$\ueta_h^{(A)}$ denotes the subset of 
$\ueta_h=(\eta_{h1},\dots,\eta_{hn})^T$ that corresponds to areas $i$ for which 
$\delta_{i} \geq h$, that is areas $i$ for which $z_{hi}$ has been obtained in step $7.$ 
Lastly, $\ueta^{(D)}$ denotes the elements of $\ueta$ not in $\ueta^{(A)}$.
A sequence of three steps achieves the objective: a. $\pi(\ueta^{(A)}|\uz,\alpha,\phi,\lambda)$, 
b. $\pi(\alpha,\phi,\lambda|\ueta^{(A)})$, and 
c. $\pi(\ueta^{(D)}|\ueta^{(A)},\alpha,\phi,\lambda)$. 
     
\begin{enumerate}

\item[8.] Let $n_{h+}$ denote the number of areas for which $\delta_{i} \geq h$.
Further, let $\tilde{\uz}_h = \{z_{hi}, i : \delta_{i} \geq h\}$.

Then, $\ueta_h^{(A)}$ is imputed from  
\begin{eqnarray}
\ueta_h^{(A)} \sim N_{n_{h+}}\Big\{
\uB (\alpha \phi^2 \uSigma_{h}^{(AA)^{-1}} \mathbf1_{n_{h+}} + \text{Diag}\{n_i^{(h)}\}\tilde{\uz}_h),
\uB \equiv \big(\phi^2 \uSigma_{h}^{(AA)^{-1}} + \text{Diag}\{1,\dots,1\}\big)^{-1}\Big\},\nonumber
\end{eqnarray}
where $\uSigma_{h}^{(AA)}$ denotes the subset of $\uQ_{\lambda}^{-1}$ from which columns 
and rows that correspond to areas with $\delta_{i} < h$ have been removed. 

\item[9.] Update $(\alpha,\phi,\lambda)$ from 
\begin{eqnarray}\label{lab1}
&& f(\alpha,\phi,\lambda|\dots) \propto \pi(\alpha,\phi,\lambda) \prod_{h} 
f(\ueta_h^{(A)}|\alpha,\phi,\lambda) \propto \pi(\alpha,\phi,\lambda) (\phi^2)^{\frac{n^*}{2}} \nonumber\\
&& \times \prod_{h} |\uQ_{\lambda h}|^{\frac{1}{2}} 
\exp \Big[-(\phi^2/2) 
(\eta_h^{(A)}-\mathbf1_{n_{h+}}\alpha)^T \uQ_{\lambda h} (\eta_h^{(A)}-\mathbf1_{n_{h+}}\alpha) 
\Big],
\end{eqnarray}
where $n^*=\sum_h n_{h+}$ and $\uQ_{\lambda h} = \lambda \uA_{h} + I_{n_{h+}}$,
in which $\uA_{h}$ denotes the $n_{h+} \times n_{h+}$ submatrix of the adjacency matrix $\uA$
obtained by removing from it columns and rows that correspond to areas for which $\delta_{i} < h$.
Note that although the $i$th diagonal element of $\uA$, $i=1,\dots,n,$ represents the numbers of neighbors
of area $i$ in the original map, the $i$th diagonal element of $\uA_h$, $i=1,\dots,n_{h+}$, is larger than or 
equal to the number of neighbors of area $i$ in the corresponding reduced map.  
Thus, the quadratic form that appears in the exponent of (\ref{lab1}) is equivalent to
$\lambda
\sum_{i' \sim i}(\eta_{hi}^{(A)}-\eta_{hi'}^{(A)})^2+
\sum_{i=1}^{n_{h+}} (\eta^{(A)}_{hi}-\alpha)^2+
\lambda \sum_{i=1}^{n_{h+}} r_{hi} (\eta^{(A)}_{hi}-\alpha)^2$,
where $r_{hi}$ is the difference between the $i$th diagonal element of $\uA_h$ 
and the number of neighbors of the $i$th area, $i=1,\dots,n_{h+}$.  

Thus, with a $N(\mu_{\alpha},\sigma^2_{\alpha})$ prior for $\alpha$, we update
\begin{eqnarray}
\alpha|\dots \sim N\left(
\frac{\phi^2 n^* \bar{\eta}^{(A)}+\lambda \phi^2 \sum_{h,i}r_{hi} \eta_{hi}^{(A)}+
\sigma^{-2}_{\alpha}\mu_{\alpha}}{n^*\phi^2+\lambda \phi^2 \sum_{h,i}r_{hi}+\sigma^{-2}_{\alpha}},
\frac{1}{n^*\phi^2+\lambda \phi^2 \sum_{h,i}r_{hi}+\sigma^{-2}_{\alpha}}\right). \nonumber
\end{eqnarray}
In our analyses we take $\mu_{\alpha} = 0.0$ and $\sigma^2_{\alpha} = 1.0$.

Further, with a Gamma$(\alpha_{\phi},\beta_{\phi})$ prior on $\phi^2$, we have that 
\begin{eqnarray}
&&\phi^2|\dots \sim \text{Gamma}\Big(\alpha_{\phi} + n^* /2,\beta_{\phi} +
\frac{1}{2} \sum_{h}\Big\{\lambda
\sum_{i' \sim i}(\eta_{hi}^{(A)}-\eta_{hi'}^{(A)})^2+
\nonumber\\ &&\sum_{i=1}^{n_{h+}} (\eta_{hi}^{(A)}-\alpha)^2 
+ \lambda \sum_{i=1}^{n_{h+}} r_{hi} (\eta^{(A)}_{hi}-\alpha)^2\Big\}\Big).\nonumber
\end{eqnarray}
In our analyses we take $\alpha_{\phi} = 1.0$ and $\beta_{\phi} = 0.1$ implying a mean of ten and 
a variance of a hundred.

Lastly, with prior $\lambda \sim \text{Unif}[0,M_{\lambda}]$, a Metropolis-Hastings step is needed.
With $\lambda_c$ and $\lambda_p$ denoting the current and proposed values,
the acceptance probability is $\min(1,P)$ where
\begin{eqnarray}
&&P=I[0 < \lambda < M_{\lambda}]\prod_h \Big\{ \prod_{i=1}^{n_h}(\lambda_p e_{ih}+1)^{\frac{1}{2}} (\lambda_c e_{ih}+1)^{-\frac{1}{2}}\Big\} \nonumber\\
&& \times \exp\Big\{-\frac{\phi^2}{2}(\lambda_p-\lambda_c) \sum_{h}\Big[
\sum_{i' \sim i}(\eta_{hi}^{(A)}-\eta_{hi'}^{(A)})^2 + 
\sum_{i=1}^{n_{h+}} r_{hi} (\eta^{(A)}_{hi}-\alpha)^2\Big]\Big\}.\nonumber
\end{eqnarray}

\item[10.] 
To sample from 
$\pi(\ueta_h^{(D)}|\ueta_h^{(A)},\alpha,\phi,\lambda)$ we let 
$n_{h-}=n-n_{h+}$. We partition the covariance matrix of 
$\ueta_h = (\ueta_h^{{(A)}^T},\ueta_h^{{(D)}^T})^T$, which is $\phi^{-2} \uQ_{\lambda}^{-1}$,
as follows $\left[ 
\begin{array}{cc}
\uSigma_h^{(AA)} & \uSigma_h^{(AD)} \\ 
\uSigma_h^{(DA)} & \uSigma_h^{(DD)}
\end{array}
\right]$. It can be seen that sampling from
$\pi(\ueta_h^{(D)}|\ueta_h^{(A)},\alpha,\phi,\lambda)$ is equivalent 
to sampling from 
$N_{n_{h-}}(\ueta_h^{(D)};\umu_{h}^{(D|A)},\uSigma_{h}^{(D|A)})$, where 
\begin{eqnarray}
\umu_{h}^{(D|A)} = \alpha \mathbf{1}_{n_{h-}} + \uSigma_h^{(DA)} \uSigma_h^{{(AA)}^{-1}}
(\ueta_h^{(A)}-\alpha  \mathbf{1}_{n_{h+}}) \text{\;and\;}
\uSigma_{h}^{(D|A)} =\uSigma_h^{(DD)}-\uSigma_h^{(DA)}\uSigma_h^{{(AA)}^{-1}}\uSigma_h^{(AD)}.\nonumber
\end{eqnarray}
\end{enumerate}

\bibliographystyle{apalike2}
\bibliography{bnpsm.bib}

\end{document}